\documentclass[aps,prd,twocolumn,superscriptaddress,floatfix,10pt]{revtex4-2}
\usepackage{xcolor}
\usepackage{graphicx}
\usepackage{subcaption}
\usepackage{amsfonts}
\usepackage{amsmath}
\usepackage{amssymb}
\usepackage{hyperref}
\usepackage[all]{hypcap}
\usepackage{cleveref}
\usepackage[T1]{fontenc}

\newcommand{\psub}[1]{{\bf #1:} }
\renewcommand{\psub}[1]{}

\definecolor{es-blue}{rgb}{0,0.4,0.8}

\hypersetup{
colorlinks=true,
linkcolor=blue,
citecolor=blue,
urlcolor=blue
}
\usepackage[
        retain-explicit-plus=true,
        retain-unity-mantissa=false,
        range-phrase=--,
        range-units=single
    ]{siunitx}
\graphicspath{{figures/}}

\newcommand{\LLOmeanSQZ}{2.23} 
\newcommand{\LLOstdSQZ}{0.36} 

\begin{document}

\title{Machine Learning for Quantum-Enhanced Gravitational-Wave Observatories}

\author{Chris Whittle}
\affiliation{LIGO, Massachusetts Institute of Technology, Cambridge, MA 02139, USA}
\affiliation{Institute of Artificial Intelligence and Fundamental Interaction, Massachusetts Institute of Technology, Cambridge, MA 02139, USA}

\author{Ge Yang}
\affiliation{Institute of Artificial Intelligence and Fundamental Interaction, Massachusetts Institute of Technology, Cambridge, MA 02139, USA}

\author{Matthew Evans}
\affiliation{LIGO, Massachusetts Institute of Technology, Cambridge, MA 02139, USA}

\author{Lisa Barsotti}
\affiliation{LIGO, Massachusetts Institute of Technology, Cambridge, MA 02139, USA}
\affiliation{Institute of Artificial Intelligence and Fundamental Interaction, Massachusetts Institute of Technology, Cambridge, MA 02139, USA}

\date{\today}

\begin{abstract}
Machine learning has become an effective tool for processing the extensive data sets produced by large physics experiments. Gravitational-wave detectors are now listening to the universe with quantum-enhanced sensitivity, accomplished with the injection of squeezed vacuum states.
Squeezed state preparation and injection is operationally complicated, as well as highly sensitive to environmental fluctuations and variations in the interferometer state.
Achieving and maintaining optimal squeezing levels is a challenging problem and will require development of new techniques to reach the lofty targets set by design goals for future observing runs and next-generation detectors.
We use machine learning techniques to predict the squeezing level during the third observing run of the Laser Interferometer Gravitational-Wave Observatory (LIGO) based on auxiliary data streams, and offer interpretations of our models to identify and quantify salient sources of squeezing degradation.
The development of these techniques lays the groundwork for future efforts to optimize squeezed state injection in gravitational-wave detectors, with the goal of enabling closed-loop control of the squeezer subsystem by an agent based on machine learning.
\end{abstract}

\maketitle


\section{Introduction}
\label{sec:intro}
\psub{Use of machine learning in gravitational-wave physics and experimental physics}
Machine learning has seen an abundance of applications in experimental physics: active control of high energy particle experiments~\cite{Kain2020}, quantum state tomography~\cite{Torlai2018,Hsieh22}, Bose-Einstein condensate preparation~\cite{Vendeiro2022} and adaptive feedback for phase estimation~\cite{Hentschel10}, to name some recent uses.
Laser interferometric gravitational-wave detectors like LIGO~\cite{aLIGOdesign}, Virgo~\cite{VirgoDesign}, GEO600~\cite{Dooley_2016} and KAGRA~\cite{Akutsu2019} are complex instruments that require hundreds of control loops to operate simultaneously. The incorporation of machine learning into the operation of these detectors is an active area of research, following the successful application of machine learning in gravitational-wave data characterization and analysis, including glitch classification~\cite{PhysRevD.88.062003,Powell_2015,Zevin_2017,PhysRevD.95.104059}, noise subtraction~\cite{Ormiston2020,Vajente2020,Hang2022}, waveform modeling~\cite{Doctor2017} and signal searches~\cite{Baker2015}.
A first demonstration of neural network-based alignment of the GEO600 optical cavities was recently performed~\cite{mukund2023demonstration}. 

Here we investigate the application of machine learning to minimize quantum noise in gravitational-wave detectors by optimizing the injection of squeezed states.

\psub{Role of squeezing in gravitational-wave detectors}
The third observing run (O3) of the gravitational-wave detector network saw the first demonstration of quantum-enhanced gravitational-wave detection~\cite{Tse2019,Acernese2019,Lough2021}. At high
(\qty{\gtrsim 50}{\hertz})
frequencies, interferometers are limited by quantum shot noise due to the random arrival times of the uncorrelated photons that make up the electromagnetic field striking the readout photodiodes.
Squeezed states can circumvent this limit by inducing correlations between photons, thereby reducing uncertainty in the readout quadrature~\cite{Barsotti_2018}. Squeezed vacuum states enabled a reduction of quantum noise by up to \qty{3.2}{\dB} in LIGO~\cite{Tse2019} and Virgo~\cite{Acernese2019}, and up to \qty{6}{\dB} in GEO600~\cite{Lough2021}.

\psub{Future demands for high levels of squeezing}
Although already a triumph in quantum metrology and astrophysics, future gravitational-wave detectors with even greater sensitivity require further improvements.
Inspection of the detector data recorded over the course of O3 reveals that it is difficult to maintain optimal squeezing performance throughout the 1-year run, with an average observed squeezing of \qty{\LLOmeanSQZ}{dB}, nearly \qty{1}{dB} less than the maximum observed.
Moreover, there is significant variability in the observed squeezing level.
For example, the histogram of squeezing levels observed in the Livingston detector (\Cref{fig:schematic}c) shows a \qty{\LLOstdSQZ}{\dB} standard deviation in the distribution of squeezing levels measured throughout the run.
Similar performance is reported by the Virgo (Fig.~3 in Ref.~\cite{Acernese2019}) and GEO600 (Fig.~2 in Ref.~\cite{Lough2021}) detectors.
Indeed, squeezed vacuum states are highly sensitive to a number of parameters of the squeezer subsystem and the interferometer: most prominently the temperature of, and pump power circulating within, the optical parametric oscillator (OPO), the thermal state of the interferometer, and the alignment of the optics directing the squeezed state into the interferometer.
Beyond these direct dependencies, other fluctuations can affect one of many control systems---power in the locking field or magnitude of the local ground motion for instance---and, in turn, the squeezing level realized in the interferometer.
Imminently, optical filter cavities are being installed in existing facilities~\cite{McCuller2020,Zhao2020,Akutsu2019} to provide a reduction in quantum noise across the entire frequency band, which further increases the complexity of the squeezed state manipulation and injection.
On the more distant horizon, next-generation gravitational-wave detectors demand squeezing levels up to \qty{10}{\dB}~\cite{CEHS, ET2020}.
Extrapolating the effect of current degradation mechanisms to the high squeezing levels of future detectors: the fluctuation in optical loss required to explain the O3 squeezing level variation (for the middle \qty{90}{\percent} of squeezing levels; \Cref{fig:schematic}c) would cause a squeezing source capable of \qty{10}{\dB} of squeezing to deteriorate to below \qty{6}{\dB}.
Consistently operating at the desired amount of squeezing in the face of increasingly complex detectors will require new methods for optimizing and tuning the injection of squeezed states throughout observing runs.


\psub{Purpose of this paper}
In the work described in this paper we trained neural network models to predict the squeezing level in the interferometer using various squeezer and witness channels---which monitor the physical environment of the interferometer---as input.
The models were trained on historical data recorded at the LIGO Livingston Observatory (LLO) during O3.
The exact inputs to the models were selected using a genetic algorithm, which additionally allowed us to identify and rank the statistical correlations present between each of the monitor channels and the squeezing level.
We performed a sensitivity analysis on the resulting model to further study correlations with witness channels and identify avenues through which the squeezer might be optimized.
This work is intended to lay the foundation for future closed-loop control of squeezer systems using machine learning by demonstrating the sensing side of the machine learning feedback loop.
Future work will close the loop by actuating on the squeezer subsystem and interferometer to implement the optimizations we identify here.

\psub{Paper contents}
In \Cref{sec:setup} we describe the experimental setup of the LIGO interferometer and the squeezed state generation.
We lay out and motivate the particular model architecture used in \Cref{sec:ml}, and outline the specific data channels we use as inputs to our model (also in \Cref{app:channels}).
\Cref{sec:analysis} describes
our analysis of the generated models and presents some findings on the behavior of the LIGO squeezer.
A discussion of a roadmap for the future is offered in \Cref{sec:future}.

\section{Experimental Setup}
\label{sec:setup}
The LIGO detector is a Michelson-based laser interferometer with \qty{4}{\kilo\metre} long arms which transduces 
space-time perturbations produced by the passage of gravitational waves into electrical signals at the interferometer readout.
The mirrors reside inside an ultra-high-vacuum system, and are isolated from the ground by multi-stage pendulums mounted on a seismically isolated platform. 
The relative distance and alignment of the interferometer's mirrors are sensed by photodetectors and wavefront sensors respectively, and actively controlled by digital servo systems that feed control signals back to actuators which steer the mirrors' relative longitudinal positions and orientations.

\begin{figure*}
    \centering
    \includegraphics[width=\hsize]{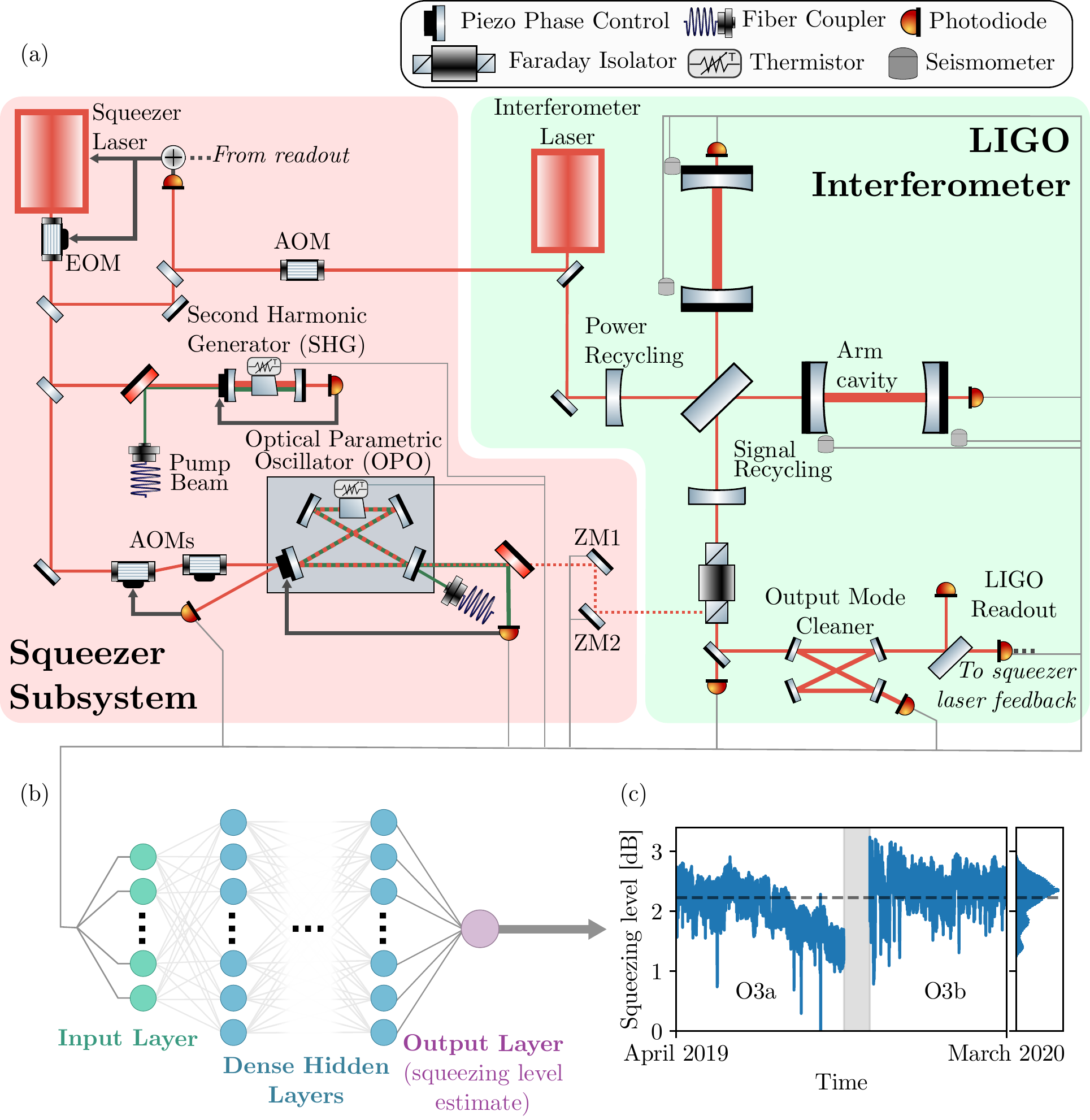}
    \caption{
    (a)
    High-level schematic of the Advanced LIGO squeezer subsystem injection into the main interferometer.
    Control sidebands are imprinted onto the squeezer laser using acousto-/electro-optic modulators (A/EOMs).
    These sidebands are used to attain phase stability of the second harmonic generator (SHG)---used to generate the green pump field---and the optical parametric oscillator (OPO), which produces the squeezed vacuum state.
    Thermoelectric coolers keep each crystal at a steady temperature.
    Steering mirrors (ZM1 and ZM2) direct the squeezed vacuum into the interferometer.
    Gray lines trace paths from readbacks from the squeezer feedback loops and representative witness sensors to the input of the machine learning model.
    Included among these auxiliary channels are data produced by photodiodes (both power measurements and alignment error signals), seismometers and thermistors.
    (b)
    The network is comprised of a series of hidden dense layers with which it calculates an estimate of the squeezing level observed in the interferometer.
    The full list of channels used as input is given in Table~\ref{tab:channels}.
    (c)
    A time series of the squeezing level observed throughout O3, measured using the cross-correlation of the interferometer readout photodiodes, also plotted as a histogram.
    The dashed line denotes the mean squeezing level of \qty{\LLOmeanSQZ}{\dB} throughout the run.
    }
    \label{fig:schematic}
\end{figure*}

Photon shot noise is a fundamental limit of the interferometer that arises from statistical fluctuations in the photon arrival time at the interferometer's readout port, and it scales inversely with the square root of the circulating power.
Nearly \qty{250}{\kilo\watt} of laser light was circulating in the LIGO arm cavities during O3, requiring a thermal compensation system to compensate for the thermal lensing produced by heating from the laser beam.
In addition to maximizing the amount of circulating laser light in the interferometer arms, the injection of squeezed vacuum states is a complementary method to further reduce shot noise.
A simplified scheme of the LIGO squeezed vacuum system is shown in \Cref{fig:schematic}a, with a full description available in~\cite{Tse2019}.
It is built around a bow-tie OPO placed on a seismically isolated platform within the ultra-high-vacuum envelope.
A \qty{1064}{\nano\metre} wavelength laser (``Squeezer Laser'' in \Cref{fig:schematic}) is phase-locked to the interferometer laser operating at the same wavelength.
From this laser, a pump \qty{532}{\nano\metre} field is produced by a second harmonic generator (SHG) and injected via optical fiber into the OPO cavity.
A \qty{1064}{\nano\metre} squeezed field is produced by the second-order nonlinearity in the periodically poled potassium titanyl phosphate (PPKTP) crystal contained within the OPO.
The squeezing level of the vacuum state produced by the OPO is determined by the amount of pump light injected and by the temperature of the non-linear crystal, both of which are under feedback control.

The resulting squeezed vacuum field is steered via two suspended mirrors (``ZMs'')~\cite{Slagmolen2011.RSI} to the output Faraday isolator which delivers the field \textit{into} the readout port of the interferometer.
An auxiliary control field consisting of a single radio-frequency sideband (at \qty{+3.1}{MHz} relative to the main laser frequency, generated by two acousto-optic modulators in series) is injected via a second optical fiber into the OPO as a proxy for sensing the phase of the squeezed vacuum field, and is additionally used to control this phase~\cite{Vahlbruch2006a, Chelkowski2007}.

In full operation, the control field co-propagates with the squeezed field, and its beat note with the interferometer field is sensed at the readout photodetectors in transmission of the output mode cleaner (OMC)~\cite{GEOlongterm}.
This signal is then fed back to the squeezer laser to lock the squeezed quadrature angle.
The relative angular alignment between the squeezed and interferometer fields is sensed by wavefront sensors placed at a pickoff before the OMC~\cite{Schreiber:16}, with alignment signals generated from the beat note between the control field and a \qty{45}{\mega\hertz} sideband of the interferometer field, and fed back to the steering optics (ZM1 and ZM2 in \Cref{fig:schematic}).

With the squeezed field phase-stabilized and well-aligned to the interferometer, the noise at the interferometer readout will decrease at frequencies where photon shot noise dominates.
We estimate the level of quantum noise reduction from the two photodetector readout signals after the OMC: the difference between these channels yields the shot noise alone, the classical noises having been removed by the common mode rejection.
While the difference is unchanged by the injection of squeezing, the sum will show the reduction in noise, and we therefore build an estimate of the squeezing level using the readout sum relative to the difference.
We can improve this slightly by subtracting an estimate of the classical noise sources from the readout sum, which itself is calculated by cross-correlating the readout channels at a reference time without any squeezing.
Finally, the squeezing level is calculated within a frequency band \qtyrange{1.1}{1.9}{\kilo\hertz} where quantum noise is dominant and \mbox{classical noises are minimal~\cite{Tse2019}}.

\section{Machine Learning Model to predict squeezing levels}
\label{sec:ml}
The primary goal of this work was to build a machine-learning-based model to accurately predict the measured squeezing level using information from auxiliary data streams (``channels'').
A successful model must capture the squeezing degradation mechanisms described in \Cref{sec:setup} and should illuminate potentially actionable correlations with the auxiliary channels.
\vspace{-0.2in}
\subsection{Inputs}\label{sec:inputs}
\vspace{-0.1in}
As input to the model, we selected channels that should capture the state of both the squeezer subsystem and the general alignment of the interferometer.
(See \Cref{tab:channels} in \Cref{app:channels} for the complete list of channels used.)

As part of the former category, we include the pump power incident on the OPO crystal (using the power reflected and transmitted from the OPO cavity as proxies), and the temperature of the OPO crystal (as controlled by the thermoelectric cooler), both of which have a direct impact on the amount of squeezing generated.
We additionally include the temperature of the SHG crystal, used to generate the OPO pump field.

We include a number of channels to characterize the interaction between the generated squeezed state and the interferometer.
The pitch and yaw of the two suspended squeezed state relay optics are the most direct signals that track this alignment.
Other signals used for general alignment sensing and control (ASC) of the full interferometer are also included.
We also use channels that track the amount of thermal compensation applied to the test masses as a proxy of the thermal state of the interferometer.
Finally, data from seismometers tracking microseismic motion at various frequency bands from 0.1 to \qty{1}{\hertz} are also included in our analysis.

We used data from all of these channels over the course of O3 to train and validate our model.
As we are focused here on long-term drifts in performance of the squeezer, we take the mean value of each channel over a minute timescale.
For some alignment channels, we also include the root mean square (RMS) values: the average will show long-term alignment drifts while the RMS shows alignment noise.
Times that coincide with large transients in the interferometer are cut from our pool of data, identified by a short-term elevation of noise or inconsistency in squeezing estimated from different frequency bands.
\vspace{-0.2in}
\subsection{Architecture}
\label{sec:architecture}
Related previous works have used linear techniques, such as coherence calculations or lasso regression~\cite{Walker2018}, to identify correlations between auxiliary channels and the astrophysical detector range.
While the efficacy of these models has been demonstrated, a more effective model of the squeezer must be capable of fitting the nonlinearities associated with squeezed state generation and alignment-dependent mode-matching losses~\cite{Collett1984,McCuller2021}.
Compared to simply fitting to an analytical model, machine learning techniques also allow us to identify and fit to new, unexpected dependencies.
To this end, we choose to use neural networks comprised of a series of hidden internal layers with nonlinear activation functions.

After the initial input layer, we additionally incorporate a random Fourier features (RFF) layer, mapping input features to random Fourier modes along random lines in parameter space.
Such RFF layers allow a network to uniformly sample the Fourier space of a function, thereby allowing one to circumvent the spectral bias of standard neural networks toward low frequency features~\cite{Rahimi2007,Tancik2020}.
This technique is commonly used in spatial machine learning applications like image regression and three-dimensional object regression, and in our system it ensures we capture both fine and coarse dependencies of our squeezing level on other parameters.
For example, we aim to capture both small misalignments about the optimum as well as how the squeezing level changes when optics are highly misaligned, the former of which would be typically underfitted by standard neural networks, but is a more common scenario than the latter.
Further, use of such a layer confers a practical benefit by generally offering much faster convergence when training the model.

\begin{figure}
    \centering
    \includegraphics[width=\hsize]{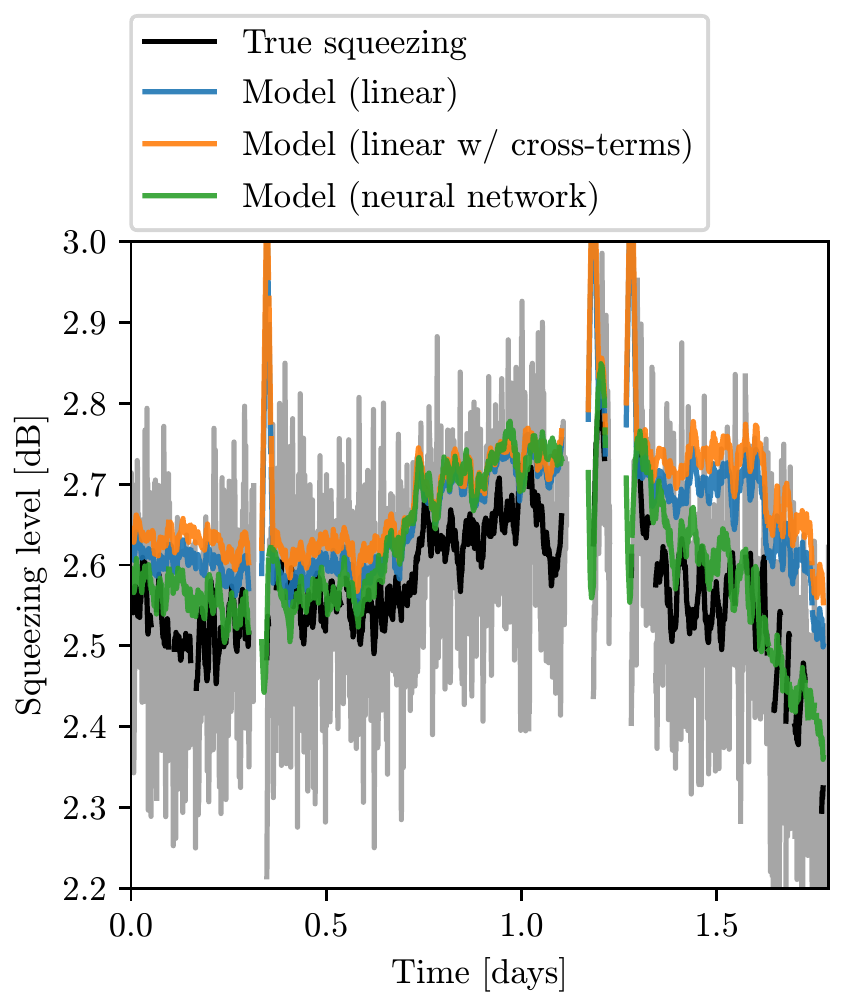}
    \caption{
    Measured squeezing level calculated from cross-correlation at the interferometer readout compared with model estimates based on squeezer, alignment and environment auxiliary channels.
    The true and estimated squeezing levels are plotted for a validation dataset with a duration of just under two days.
    Although each datapoint is a \qty{60}{\s} average, we smooth the data with a rolling window of 20 minutes to aid readability and emphasize longer-term trends (the pre-smoothed true squeezing level is plotted in gray).
    The models are all trained on the 30 days of data preceding this validation dataset.
    The full nonlinear neural network model (green) is compared with two more simple models: a linear combination of witness channels (blue) and a linear model that also allows for cross-terms (orange).  The gaps in data correspond with periods for which the interferometer was either unlocked or not squeezing, or other excess noise resulted in a veto from our data analysis pipeline.
    }
    \label{fig:time_series}
\end{figure}

A series of nonlinear dense layers follow this mapping, before the network outputs the estimated squeezing level.
We conducted a brute-force search of the hyperparameter space to empirically identify a successfull architecture by first selecting several representative time intervals from each half of O3 (i.e. O3a and O3b).
A model was trained with the given hyperparameters and evaluated using separate validation data.
Our search space spanned different activation functions, various counts of sequential dense layers, different dimensions of said dense layers, and the dimensionality (and presence) of the RFF layer.

We additionally tested different durations of training data; longer training intervals increase the amount of training data available but sacrifice accuracy as the interferometer undergoes long-term drifts that may not be captured by the auxiliary witness channels.
By using a training data duration that is shorter than the characteristic timescale of these long-term drifts, we can still achieve a model that can accurately predict the squeezing level in the short term.
We ultimately found that one month of training struck the balance between sufficient training data and model specificity.
The optimal architecture was found to use two 64-dimensional dense layers with sigmoid activation functions fitting to 4096 Fourier features output from the RFF layer.

The model performance is demonstrated in \Cref{fig:time_series}, where it is shown predicting the squeezing level over approximately one week of O3b immediately following one month of training.
The model is compared against a simple linear model, as well as a model that additionally allows for cross-terms between different witness channels.
While some segments of O3 show similar performance for each of these models, there is a significant fraction of times that do benefit from capturing the nonlinear behavior to achieve better predictions, as shown in \Cref{fig:time_series}.

During our hyperparameter optimization, we additionally tested architectures that are well-suited to time series data, including long short-term memory (LSTM) networks and one-dimensional convolutional neural networks (1D CNN).
We ultimately discovered that these alternative architectures granted marginal improvements in accuracy (\qty{\lesssim 5}{\percent} lower average loss) at the cost of a loss of interpretability.
Future work could further investigate time series models, for example through use of those that specialize in interpretability of dependencies over variable timescales~\cite{Lim2021}.
\vspace{-0.2in}
\subsection{Feature Selection}
\label{sec:feature}
\vspace{-0.1in}
One of the primary goals of this study was to identify interpretable and actionable dependencies of the squeezing level on other parameters of the interferometer.
To aid with this interpretability, as well as assuage some of the other issues with high-dimensional regression like over-fitting or lengthy training processes, we performed feature selection to identify the most predictive channels among our chosen inputs, discarding the rest.
This was accomplished with a genetic algorithm, used to select an optimal subset of channels as inputs for our model~\cite{Xue16}.
This approach benefits from being scalable and system-agnostic, but can be computationally intensive.

\begin{figure}
    \centering
    \includegraphics{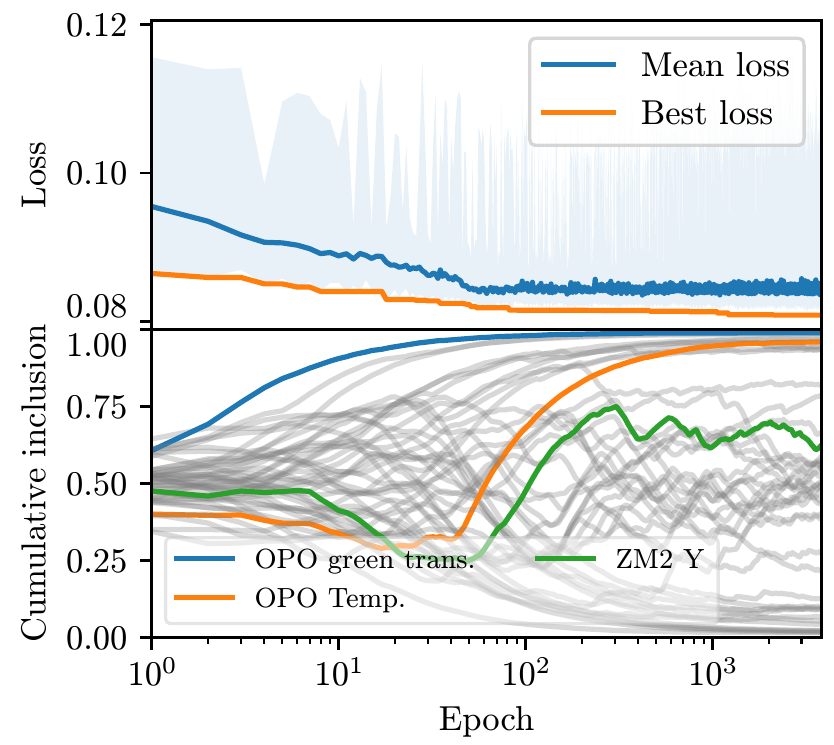}
    \caption{
    Results from feature selection of the auxiliary interferometer channels using a genetic algorithm.
    The average loss (model error, blue line) within each generation is plotted as a function of epoch number, along with the lowest loss (orange line) model trained so far (\emph{top}).
    To compare the relative importance placed on each channel by the algorithm, we also plot the fraction of models in which each channel is included, cumulatively averaged for each generation over its ancestors (\emph{bottom}).
    We highlight channels on which we expect the squeezing level to depend: the laser power and temperature within the optical parametric oscillator, as well as the yaw of the second steering mirror (ZM2) between the squeezer and interferometer.
    }
    \label{fig:genetic}
\end{figure}

We seeded our initial populations with random subsets of the channels listed in \Cref{tab:channels}.
During each iteration, models were trained on data from their respective channels over a selection of time intervals throughout O3.
The average error of each model was computed from subsequent validation intervals, which was then used to weight the probability of that subset propagating to the following generation.
The results of the genetic algorithm are shown in \Cref{fig:genetic}: the top panel shows the average loss (blue trace) of models in each generation stochastically decreasing as more accurate models are preferentially selected.
We also kept track of the lowest loss observed at each iteration (orange trace) and ultimately used the corresponding channel subset for the subsequent analyses (marked in \Cref{tab:channels}).

\section{Sensitivity Analysis}
\label{sec:analysis}
After building a model architecture and determining a predictive set of input channels, we looked for interpretable correlations with the detector squeezing level.
The field of sensitivity analysis deals with studying how the output of a model varies with each of its inputs, as well as the relative importance of each of its features.
Here, we applied sensitivity analysis techniques to identify specific channels to which the squeezing level is most sensitive, and quantify the extent of this dependence on alignment and other interferometer channels.
\vspace{-0.2in}
\subsection{Global Sensitivity Analysis}\label{sec:global}
\vspace{-0.1in}
Global sensitivity analyses explore the behavior of the model throughout the full extent of the input feature space, providing a holistic view of its dependencies.

A byproduct of our feature selection is that we have already computed a quantitative measure of the importance of each channel for squeezing level prediction.
The average fraction of models in which each channel is included gives an indication of how that feature is being weighted.
In the lower panel of \Cref{fig:genetic}, we plot the cumulative average inclusion of each channel as the genetic algorithm evolves.
As a sanity check, we inspected channels for which we know the squeezing level has a strong dependence (see the highlighted curves): laser power and crystal temperature both appear directly in the analytical form for amount of squeezing generated by an OPO~\cite{Collett1984,Goda2005}.
We likewise know that a shift in alignment into the interferometer (from e.g. motion in one of the steering mirrors ZM) should result in a change in squeezing observed.
However, because the interferometer alignment can also drift, looking at the change of a single optic in isolation has a less clear-cut consequence for squeezing.
The inclusion fractions for all channels are given in \Cref{tab:channels}.
In addition to the channels already highlighted, we saw that some seismometer channels show strong correlation with the squeezing level, as well as a handful of alignment channels for other cavities in the interferometer.

A well-established technique in machine learning literature for studying global sensitivities is by computing Sobol indices~\cite{Sobol2001}.
This technique requires decomposing a predictive model $f(\mathbf{x})$, which takes $n$-dimensional inputs $\mathbf{x}$, into summands of different dimensions,
\begin{multline}
    f(\mathbf{x}) = f_0 + \sum_i f_i(x_i) + \sum_{i<j} f_{ij}(x_i, x_j)\\
    + \cdots + f_{12\cdots n}(x_1,x_2,\ldots,x_n).
\end{multline}
Sobol indices are computed as variances of these constituent functions, normalized by the variance of the total model output.
In this way, each Sobol index represents the contribution of a given channel $x_i$---or the contribution from its interaction with one or more other channels---to the output.
The first-order indices $S_1^i$ encapsulate variations due to fluctuations in just that channel $x_i$, while the second-order indices $S_2^{i,j}$ capture variability due to the joint variation of two channels.
These effects, in addition to higher-dimension components, sum together to yield the total Sobol indices $S_T^i$.

\begin{figure*}
    \centering
    \includegraphics{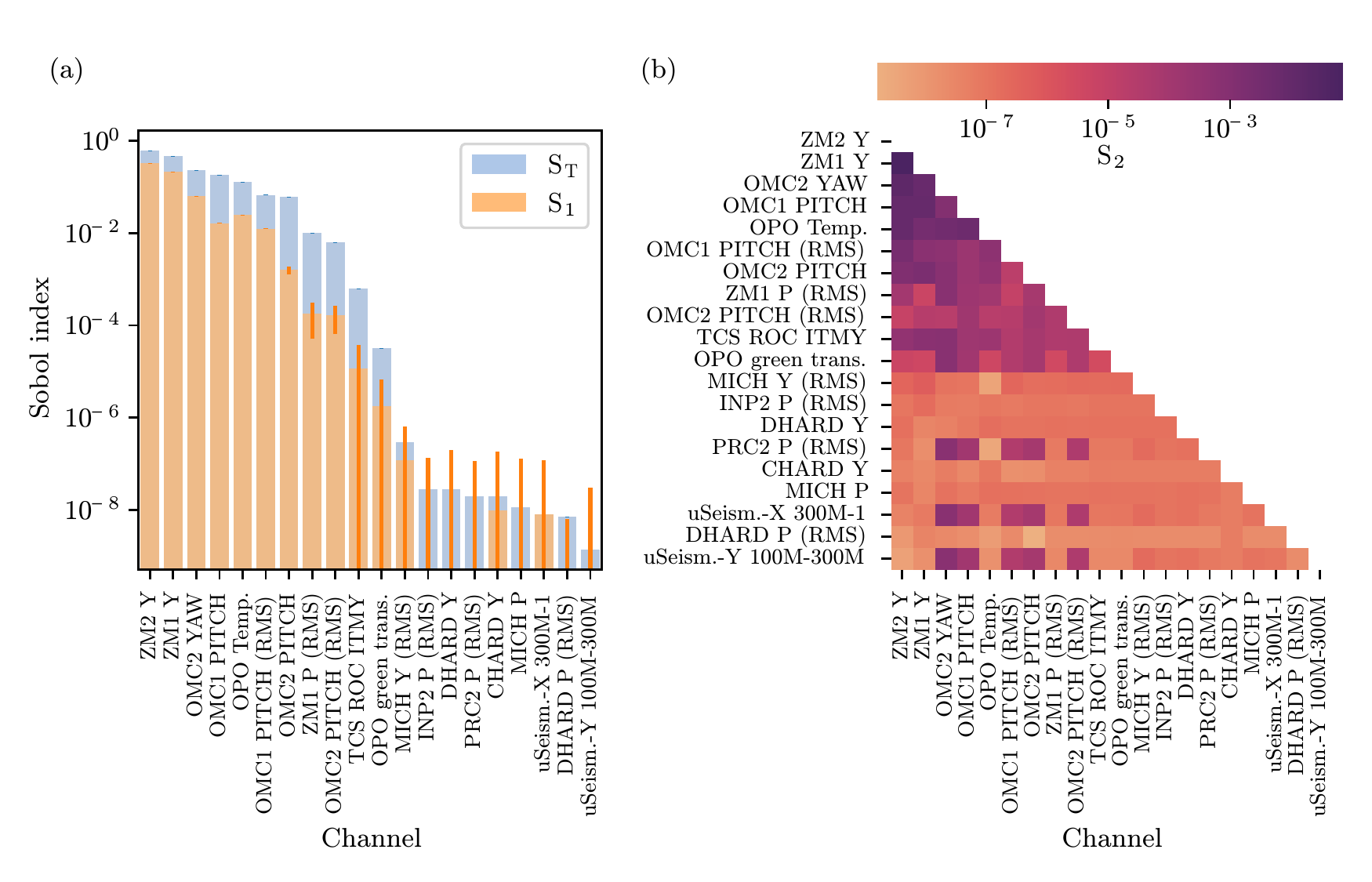}
    \caption{
        (a) First-order and total Sobol indices, showing the global dependence of the squeezing model on each channel when varied independently (first-order) and jointly with other channels (total) respectively, computed by Monte Carlo sampling.
        The orange error bars show the 95\% confidence interval for the first-order indices.
        (b) Second-order Sobol indices, showing the global dependence of the squeezing model when two channels are varied simultaneously.
    }
    \label{fig:sobol}
\end{figure*}

We trained models on four sets of month-long training data, then used Monte Carlo sampling over each of their respective input spaces to compute the variances of the decomposed model outputs.
We subsequently evaluated the Sobol indices associated with each input channel and average over each model to produce a plot of first-, second- and total-order indices, shown in \Cref{fig:sobol}.
The computational cost associated with sampling this high-dimensional input space demands the dimensionality reduction we performed through feature selection in \Cref{sec:feature}---more than a handful of additional channels would render this approach infeasible.

The first- and total-order sensitivities highlight a particular sensitivity of the model to drifts in the yaw of the steering mirrors (ZM1 and ZM2).
Fluctuations in pitch, which manifest through the RMS channels, of the first steering mirror and the output mode cleaner (OMC) optics are also shown to be significant, suggesting that improvements in the OMC pitch alignment loops could lead to more stable squeezing levels.
The second-order indices seem to mostly track the sensitivities depicted by the first- and total-order indices.

\subsection{Local Sensitivity Analysis}
\label{sec:local}
\begin{figure}
    \centering
    \includegraphics[width=\hsize]{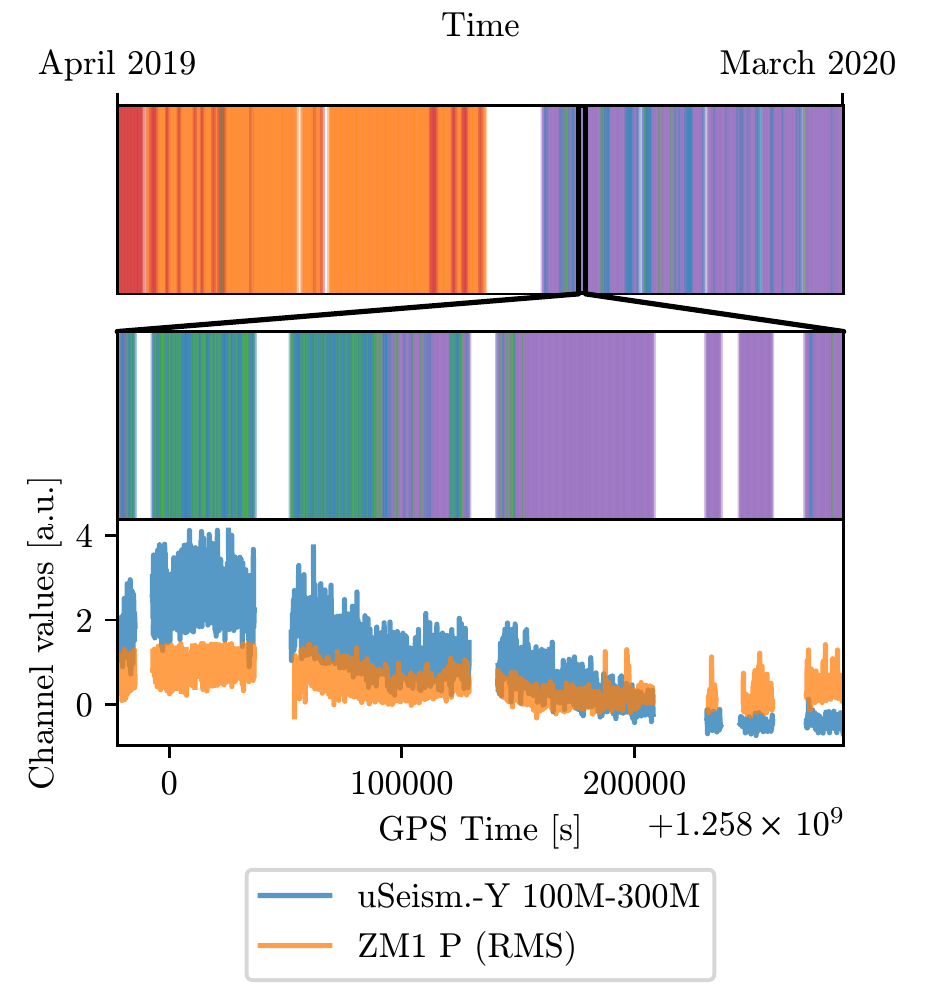}
    \caption{
        The segmentation of O3 data into $k$-means clusters, where each of the five colors represent a quantitatively distinct operating point.
        The zoomed-in segment demonstrates heightened levels of ground motion (visible in the seismometer data and in alignment error signals) corresponding to a different cluster compared with nominal low-noise operation.
    }
    \label{fig:cluster}
\end{figure}

We studied how the sensitivities of the models vary as we move to different points in parameter space.
To identify different operational regimes occupied throughout O3, we performed a $k$-means clustering over all of the input channels (\Cref{fig:cluster}).
We partitioned time samples into five different clusters---chosen such that each cluster is qualitatively distinct and can be intuitively understood.

\begin{figure*}
    \centering
    \begin{subfigure}[b]{0.49\hsize}
        \centering
        \includegraphics[width=\hsize]{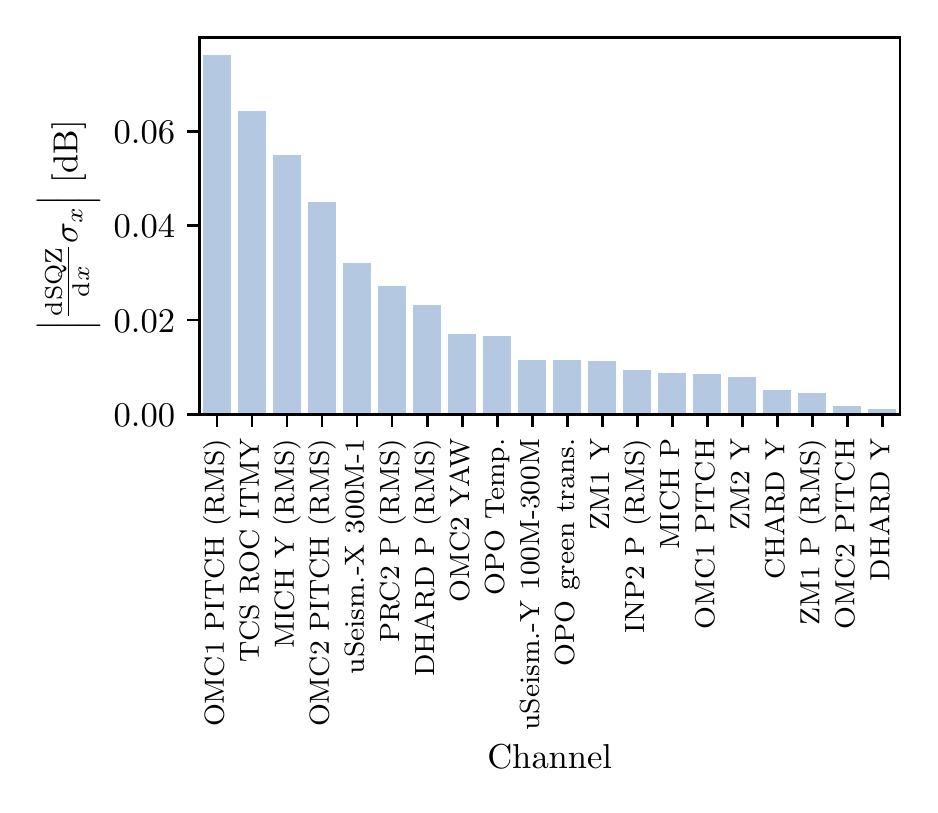}
        \caption{
            The model gradients computed at the cluster centroid representative of typical operation during O3b (shown in \Cref{fig:cluster} in purple).
        }
        \label{fig:cluster1_gradients}
    \end{subfigure}
    \begin{subfigure}[b]{0.49\hsize}
        \centering
        \includegraphics[width=\hsize]{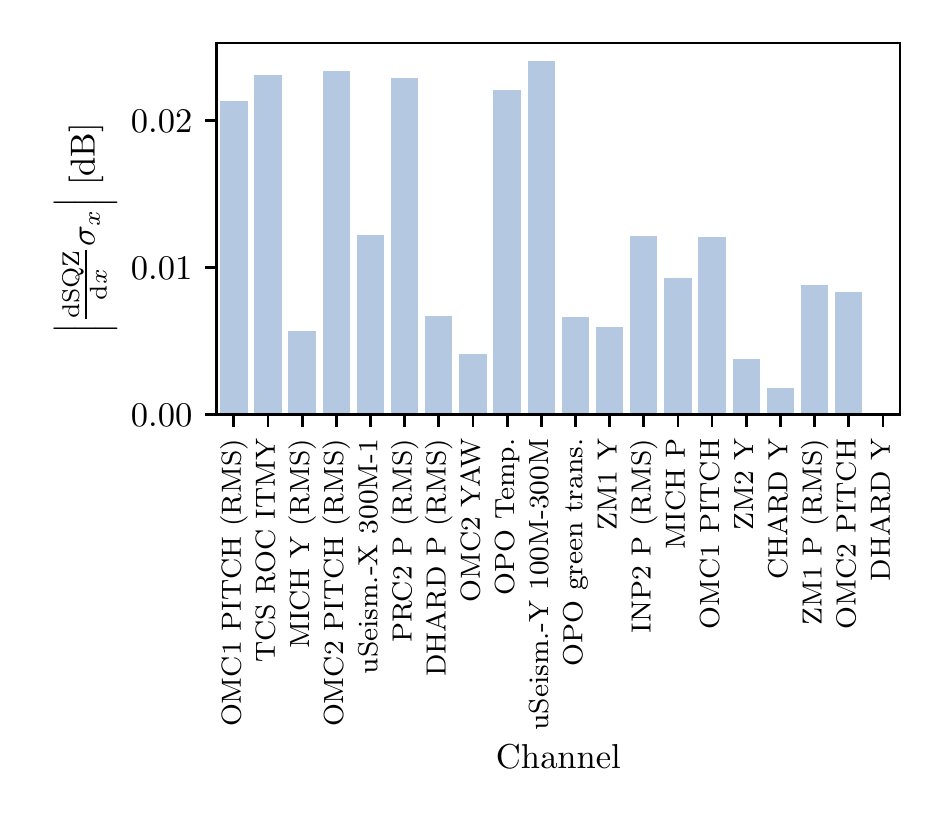}
        \caption{
            The model gradients computed at the cluster centroid characterized by higher-than-usual micro-seismic activity and alignment noise (shown in \Cref{fig:cluster} in blue).
        }
        \label{fig:cluster5_gradients}
    \end{subfigure}
    \caption{
        Partial derivatives shown at different points in parameter space for a neural network trained on \qty{\sim 1}{month} of O3b data.
        The derivative of the predicted squeezing level with respect to each auxiliary channel $\partial \mathrm{SQZ}/ \partial x$ is shown.
        Each derivative is multiplied by the auxiliary channel standard deviation $\sigma_x$, giving the squeezing level variations that result from typical fluctuations in the given channel.
        The channels are listed in order of decreasing sensitivity, according to the partial derivatives, during nominal O3b operation.
        This analysis hints at the variables on which one should actuate to optimize the squeezing level in the interferometer in a particular regime.
    }
    \label{fig:gradient}
\end{figure*}

The partial derivative of the network output can be computed with respect to each of the input factors and evaluated at particular fixed points.
We trained models over various time intervals, as in \Cref{sec:global}, and computed partial derivatives at each of the cluster centroids calculated calculated from $k$-means.
We then compared the magnitudes of the partial derivatives, after normalizing by the standard deviation of each input channel, to evaluate sensitivity to each of the input parameters.
\Cref{fig:gradient} shows this procedure for a month of training data in O3b, evaluated at a typical O3b interferometer configuration (\Cref{fig:gradient}a) and in a regime with significant micro-seismic activity and alignment noise (\Cref{fig:gradient}b).
We see that the relative effects of fluctuations in various alignment signals change during times with substantial ground motion.
We also observe that some channels to which the squeezing level is globally sensitive (shown in \Cref{fig:sobol}) are locally insensitive at particular points in parameter space.

In principle, the derivatives computed can be used to determine actions that can be taken to optimize squeezing.
For instance, a non-zero derivative in the OPO temperature suggests that it is sub-optimal, and the sign of the derivative will indicate the direction in which it should be adjusted.
Note, however, that this analysis ignores interactions between channels, which can somewhat muddy direct interpretation of these local sensitivity indices (and, in particular, their signs).

\section{Roadmap for the future}
\label{sec:future}



The work described here demonstrated that machine learning models are capable of sensing the state of the squeezer and interferometer (\Cref{sec:ml}), and can produced actionable diagnostic information about sources of squeezing degradation (\Cref{sec:analysis}).
A fruitful direction for future work is toward continuous optimization of the squeezer based on these real-time machine-learning-based prognoses.
An artificial intelligent (AI) agent that replaces \textit{ad hoc} and infrequent manual tuning of the squeezing system with a controller that is \textit{always on} may be a critical step towards achieving the \qty{10}{\dB} suppression of quantum noise required by the next-generation gravitational wave observatories~\cite{CEHS,ET2020}.
This offers a fertile ground for importing ideas from other field, and the development of new control methods.




Capturing all of the complexity of gravitational-wave interferometers in a model can be challenging, which makes applying analytical control methods such as model-predictive control difficult.
An alternative is to adopt learning based-approaches, such as model-free reinforcement learning (RL), that are able to produce control behaviors similar to those from the human operators by learning via trial-and-error in a physics simulator.
``Learning-to-control'' schemes as such have so far achieved impressive results on complex games such as the ATARI arcade learning environments (ALE, see~\citealt{bellemare2013arcade}), StarCraft, and Dota, without making explicit assumptions on the underlying decision processes~\citep{Mnih2015dqn,Vinyals2019AlphaStar,berner2019dota}.
This also comes with the added benefit that the AI agent can continuously improve in performance during deployment, given additional supervision in the form of reinforcement.

Past learning-based control methods struggle with learning and operating under confounding signals.
The feature analysis in Section~\ref{sec:feature} helps by removing confounding signals through an information-theoretic approach that has been shown to improve learning on complex learning tasks~\cite{fu2021tia}.
Recent results in sim-to-real transfer approaches, which show promise in robotics for implicitly identifying system parameters under partial observability~\cite{Kumar2021RMA,lee2020quadrupedal}, may also be applicable to gravitational-wave detectors.

\begin{figure*}
    \centering
    \includegraphics[width = 0.8\linewidth]{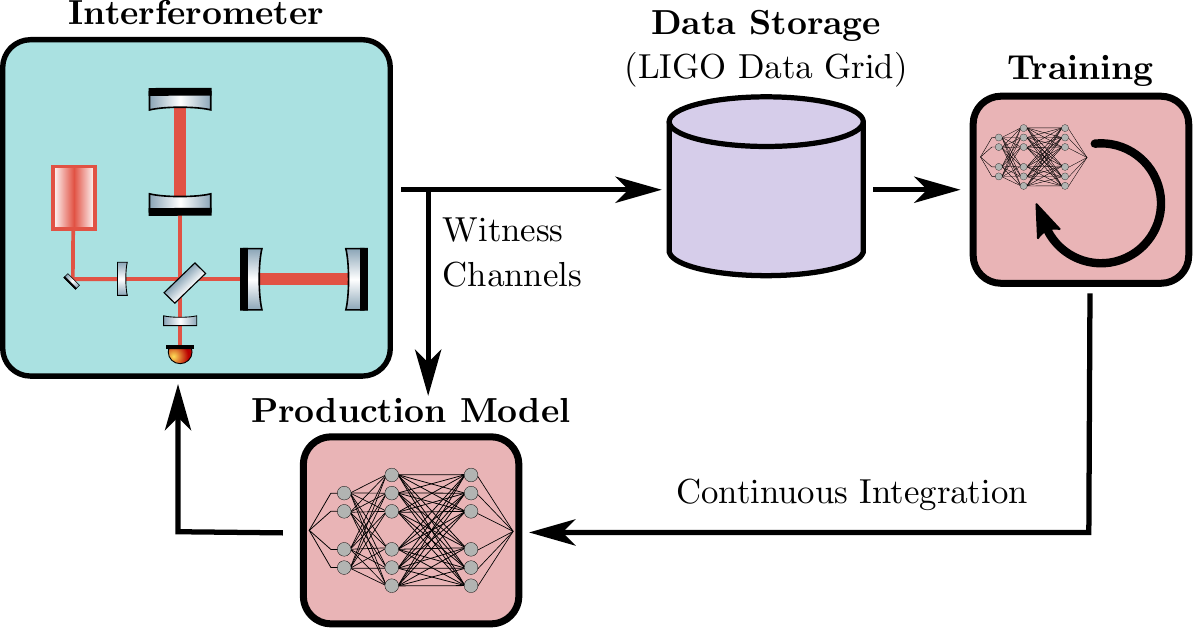}
    \caption{
        Proposed workflow for optimization of the squeezer (or other) subsystem with machine learning.
        In this scenario, the latest trained model continuously runs on the interferometer data as it is recorded, identifying possible optimizations and feeding back to the LIGO real-time control system.
        Meanwhile, additional hardware runs training on the latest data, with the updated model weights continuously deployed to the production model.
    }
    \label{fig:architecture}
\end{figure*}

Any move towards machine-learning techniques and learning-to-control will also necessitate next-generation infrastructure in the data acquisition, storage, analysis, and control pipeline~\cite{ErikNature}.
\Cref{fig:architecture} illustrates a simplified example architecture for a computational and inference infrastructure capable of running a learned controller.

\section{Conclusions}
\label{sec:conclusions}


The application of machine learning in experimental physics is growing, and it is providing new methods to optimize performance.
Here we describe an exploration of trained neural network models that estimate the quantum-enhanced performance of the LIGO Livingston detector via squeezing injection, using historical data from the past observing run O3. We have demonstrated the sensing, processing and interpretation steps of our training models, including various sensitivity analysis techniques to identify correlations between auxiliary channels and the squeezing observed in the interferometer.
Specific alignment signals for the squeezed state steering mirrors and other cavities within the interferometer have been found to be the most correlated. The most computationally expensive processes in our pipeline are the hyperparameter optimization, feature selection steps and, to a lesser extent, computation of Sobol indices.
Once a network architecture has been selected, training, evaluation and local sensitivity analysis is relatively lightweight, making the approaches demonstrated here inexpensive.

Our analysis shows that machine learning could be used to continuously monitor the performance of the squeezer system and the interferometer in a high-dimensional parameter space to identify causes of squeezing degradation.
Such continuous monitoring of squeezing degradation sources is presently not possible in existing gravitational-wave detectors, and would be a significant step toward the realization of persistently high levels of squeezing. The next step of this work would be to use this newly acquired information to actuate on the system and maximize squeezing levels.
This approach could be extended to other quantum precision experiments in which maintaining optimal performance for long periods of time is important.

Future gravitational-wave detectors like Cosmic Explorer~\cite{CEHS} will require high levels of squeezing (up to 10 dB) to reach their sensitivity targets.
In the next decade AI might become not just a useful tool, but an indispensable one to guarantee optimal performance over long periods of time.
Moreover, similar techniques can evolve to address many other interferometer control problems, with the potential to revolutionize the way current gravitational-wave detectors are controlled, and the way future observatories are designed.

\newpage
\begin{acknowledgments}
    LIGO was constructed by the California Institute of Technology and Massachusetts Institute of Technology with funding from the National Science Foundation, and operates under Cooperative Agreement No. PHY-1764464.
    Advanced LIGO was built under Grant No. PHY-0823459.
    This work is also supported by the National Science Foundation under Cooperative Agreement PHY-2019786 (The NSF AI Institute for Artificial Intelligence and Fundamental Interactions).
    The authors acknowledge the MIT SuperCloud and Lincoln Laboratory Supercomputing Center for providing HPC resources that have contributed to the research results reported within this paper.
    We thank Maggie Tse for the initial intuition that squeezing optimization could benefit from an AI agent, for brainstorming on a possible approach for this study, and for providing comments on the manuscript. 
    We thank Lee McCuller, Dhruva Ganapathy, Wenxuan Jia and Kevin Kuns for brainstorming in the initial phase of the project on causes of squeezing degradation.
    We thank Pulkit Agrawal for discussions on reinforcement learning, and Erik Katsavounidis, Philip Harris and Alec Gunny for discussions on deployment of machine learning-based algorithms in LIGO.
    Finally, we thank Nikhil Mukund for carefully reading the manuscript and providing comments, and Hang Yu for his comments as part of the LIGO review process.
    This paper has LIGO Document Number LIGO-P2300114.
\end{acknowledgments}


\appendix

\section{Channel List}
\label{app:channels}
The channels used as input for this analysis are listed in \Cref{tab:channels}.

\begin{table*}[h!]
    \caption{
        List of channels used as input for the squeezing level estimator, listed both by their standardized channel name as well as more readable, abbreviated versions.
        The {\tt mean} and {\tt rms} suffixes refer to the statistic computed over \qty{60}{\s} intervals.
        The average inclusion fraction of each channel during the feature selection process (\Cref{fig:genetic}) is also shown: a higher number indicates a greater importance of the channel for squeezing level estimation.
        Channels marked with a circle (\textbullet) were included in the optimal subset found during feature selection and used for subsequent sensitivity analysis (\Cref{sec:analysis}).
    }
    \begin{tabular}{lllr}
        \hline \hline
        & LIGO channel name & Description & Average inclusion fraction\\
        \hline
        & {\tt ASC-CHARD\_P\_OUTPUT.mean,m-trend} & CHARD P & 0.40\\
        & {\tt ASC-CHARD\_P\_OUTPUT.rms,m-trend} & CHARD P (RMS) & 0.07\\
       \textbullet & {\tt ASC-CHARD\_Y\_OUTPUT.mean,m-trend} & CHARD Y & 0.60\\
        & {\tt ASC-CHARD\_Y\_OUTPUT.rms,m-trend} & CHARD Y (RMS) & 0.10\\
        & {\tt ASC-DHARD\_P\_OUTPUT.mean,m-trend} & DHARD P & 0.36\\
       \textbullet & {\tt ASC-DHARD\_P\_OUTPUT.rms,m-trend} & DHARD P (RMS) & 0.40\\
       \textbullet & {\tt ASC-DHARD\_Y\_OUTPUT.mean,m-trend} & DHARD Y & 0.40\\
        & {\tt ASC-DHARD\_Y\_OUTPUT.rms,m-trend} & DHARD Y (RMS) & 0.06\\
        & {\tt ASC-INP2\_P\_OUTPUT.mean,m-trend} & INP2 P & 0.03\\
       \textbullet & {\tt ASC-INP2\_P\_OUTPUT.rms,m-trend} & INP2 P (RMS) & 0.99\\
        & {\tt ASC-INP2\_Y\_OUTPUT.mean,m-trend} & INP2 Y & 0.02\\
        & {\tt ASC-INP2\_Y\_OUTPUT.rms,m-trend} & INP2 Y (RMS) & 0.02\\
       \textbullet & {\tt ASC-MICH\_P\_OUTPUT.mean,m-trend} & MICH P & 0.93\\
        & {\tt ASC-MICH\_P\_OUTPUT.rms,m-trend} & MICH P (RMS) & 0.93\\
        & {\tt ASC-MICH\_Y\_OUTPUT.mean,m-trend} & MICH Y & 0.44\\
       \textbullet & {\tt ASC-MICH\_Y\_OUTPUT.rms,m-trend} & MICH Y (RMS) & 0.95\\
       \textbullet & {\tt OMC-ASC\_P1\_I\_OUTPUT.mean,m-trend} & OMC1 PITCH & 0.82\\
       \textbullet & {\tt OMC-ASC\_P1\_I\_OUTPUT.rms,m-trend} & OMC1 PITCH (RMS) & 0.98\\
       \textbullet & {\tt OMC-ASC\_P2\_I\_OUTPUT.mean,m-trend} & OMC2 PITCH & 0.46\\
       \textbullet & {\tt OMC-ASC\_P2\_I\_OUTPUT.rms,m-trend} & OMC2 PITCH (RMS) & 0.28\\
        & {\tt OMC-ASC\_Y1\_I\_OUTPUT.mean,m-trend} & OMC1 YAW & 0.65\\
        & {\tt OMC-ASC\_Y1\_I\_OUTPUT.rms,m-trend} & OMC1 YAW (RMS) & 0.05\\
       \textbullet & {\tt OMC-ASC\_Y2\_I\_OUTPUT.mean,m-trend} & OMC2 YAW & 0.69\\
        & {\tt OMC-ASC\_Y2\_I\_OUTPUT.rms,m-trend} & OMC2 YAW (RMS) & 0.08\\
        & {\tt ASC-PRC2\_P\_OUTPUT.mean,m-trend} & PRC2 P & 0.50\\
       \textbullet & {\tt ASC-PRC2\_P\_OUTPUT.rms,m-trend} & PRC2 P (RMS) & 0.62\\
        & {\tt ASC-PRC2\_Y\_OUTPUT.mean,m-trend} & PRC2 Y & 0.43\\
        & {\tt ASC-PRC2\_Y\_OUTPUT.rms,m-trend} & PRC2 Y (RMS) & 0.27\\
        & {\tt ISI-GND\_STS\_ETMX\_Y\_BLRMS\_100M\_300M.mean,m-trend} & uSeism.-X 100M-300M & 0.14\\
       \textbullet & {\tt ISI-GND\_STS\_ETMX\_Y\_BLRMS\_300M\_1.mean,m-trend} & uSeism.-X 300M-1 & 0.96\\
       \textbullet & {\tt ISI-GND\_STS\_ETMY\_Y\_BLRMS\_100M\_300M.mean,m-trend} & uSeism.-Y 100M-300M & 0.95\\
        & {\tt ISI-GND\_STS\_ETMY\_Y\_BLRMS\_300M\_1.mean,m-trend} & uSeism.-Y 300M-1 & 0.09\\
       \textbullet & {\tt SQZ-OPO\_IR\_LF\_OUTPUT.mean,m-trend} & OPO green trans. & 0.99\\
        & {\tt SQZ-OPO\_REFL\_DC\_POWER.mean,m-trend} & OPO green refl. & 0.02\\
       \textbullet & {\tt SQZ-OPO\_TEC\_THERMISTOR\_TEMPERATURE.mean,m-trend} & OPO Temp. & 0.95\\
        & {\tt SQZ-SHG\_TEC\_THERMISTOR\_TEMPERATURE.mean,m-trend} & SHG Temp. & 0.02\\
        & {\tt SUS-ZM1\_M1\_DAMP\_P\_OUTPUT.mean,m-trend} & ZM1 P & 0.57\\
       \textbullet & {\tt SUS-ZM1\_M1\_DAMP\_P\_OUTPUT.rms,m-trend} & ZM1 P (RMS) & 0.73\\
       \textbullet & {\tt SUS-ZM1\_M1\_DAMP\_Y\_OUTPUT.mean,m-trend} & ZM1 Y & 0.55\\
        & {\tt SUS-ZM1\_M1\_DAMP\_Y\_OUTPUT.rms,m-trend} & ZM1 Y (RMS) & 0.09\\
        & {\tt SUS-ZM2\_M1\_DAMP\_P\_OUTPUT.mean,m-trend} & ZM2 P & 0.53\\
        & {\tt SUS-ZM2\_M1\_DAMP\_P\_OUTPUT.rms,m-trend} & ZM2 P (RMS) & 0.47\\
       \textbullet & {\tt SUS-ZM2\_M1\_DAMP\_Y\_OUTPUT.mean,m-trend} & ZM2 Y & 0.69\\
        & {\tt SUS-ZM2\_M1\_DAMP\_Y\_OUTPUT.rms,m-trend} & ZM2 Y (RMS) & 0.44\\
       \textbullet & {\tt TCS-SIM\_ITMY\_SURF\_ROC\_FULL\_OUTPUT.mean,m-trend} & TCS ROC ITMY & 0.97\\
       \hline \hline  
    \end{tabular}
    \label{tab:channels}
\end{table*}

\clearpage
\bibliography{references}

\begin{thebibliography}{49}%
\makeatletter
\providecommand \@ifxundefined [1]{%
 \@ifx{#1\undefined}
}%
\providecommand \@ifnum [1]{%
 \ifnum #1\expandafter \@firstoftwo
 \else \expandafter \@secondoftwo
 \fi
}%
\providecommand \@ifx [1]{%
 \ifx #1\expandafter \@firstoftwo
 \else \expandafter \@secondoftwo
 \fi
}%
\providecommand \natexlab [1]{#1}%
\providecommand \enquote  [1]{``#1''}%
\providecommand \bibnamefont  [1]{#1}%
\providecommand \bibfnamefont [1]{#1}%
\providecommand \citenamefont [1]{#1}%
\providecommand \href@noop [0]{\@secondoftwo}%
\providecommand \href [0]{\begingroup \@sanitize@url \@href}%
\providecommand \@href[1]{\@@startlink{#1}\@@href}%
\providecommand \@@href[1]{\endgroup#1\@@endlink}%
\providecommand \@sanitize@url [0]{\catcode `\\12\catcode `\$12\catcode
  `\&12\catcode `\#12\catcode `\^12\catcode `\_12\catcode `\%12\relax}%
\providecommand \@@startlink[1]{}%
\providecommand \@@endlink[0]{}%
\providecommand \url  [0]{\begingroup\@sanitize@url \@url }%
\providecommand \@url [1]{\endgroup\@href {#1}{\urlprefix }}%
\providecommand \urlprefix  [0]{URL }%
\providecommand \Eprint [0]{\href }%
\providecommand \doibase [0]{https://doi.org/}%
\providecommand \selectlanguage [0]{\@gobble}%
\providecommand \bibinfo  [0]{\@secondoftwo}%
\providecommand \bibfield  [0]{\@secondoftwo}%
\providecommand \translation [1]{[#1]}%
\providecommand \BibitemOpen [0]{}%
\providecommand \bibitemStop [0]{}%
\providecommand \bibitemNoStop [0]{.\EOS\space}%
\providecommand \EOS [0]{\spacefactor3000\relax}%
\providecommand \BibitemShut  [1]{\csname bibitem#1\endcsname}%
\let\auto@bib@innerbib\@empty
\bibitem [{\citenamefont {Kain}\ \emph {et~al.}(2020)\citenamefont {Kain},
  \citenamefont {Hirlander}, \citenamefont {Goddard}, \citenamefont {Velotti},
  \citenamefont {Della~Porta}, \citenamefont {Bruchon},\ and\ \citenamefont
  {Valentino}}]{Kain2020}%
  \BibitemOpen
  \bibfield  {author} {\bibinfo {author} {\bibfnamefont {V.}~\bibnamefont
  {Kain}}, \bibinfo {author} {\bibfnamefont {S.}~\bibnamefont {Hirlander}},
  \bibinfo {author} {\bibfnamefont {B.}~\bibnamefont {Goddard}}, \bibinfo
  {author} {\bibfnamefont {F.~M.}\ \bibnamefont {Velotti}}, \bibinfo {author}
  {\bibfnamefont {G.~Z.}\ \bibnamefont {Della~Porta}}, \bibinfo {author}
  {\bibfnamefont {N.}~\bibnamefont {Bruchon}},\ and\ \bibinfo {author}
  {\bibfnamefont {G.}~\bibnamefont {Valentino}},\ }\href
  {https://doi.org/10.1103/PhysRevAccelBeams.23.124801} {\bibfield  {journal}
  {\bibinfo  {journal} {Phys. Rev. Accel. Beams}\ }\textbf {\bibinfo {volume}
  {23}},\ \bibinfo {pages} {124801} (\bibinfo {year} {2020})}\BibitemShut
  {NoStop}%
\bibitem [{\citenamefont {Torlai}\ \emph {et~al.}(2018)\citenamefont {Torlai},
  \citenamefont {Mazzola}, \citenamefont {Carrasquilla}, \citenamefont
  {Troyer}, \citenamefont {Melko},\ and\ \citenamefont {Carleo}}]{Torlai2018}%
  \BibitemOpen
  \bibfield  {author} {\bibinfo {author} {\bibfnamefont {G.}~\bibnamefont
  {Torlai}}, \bibinfo {author} {\bibfnamefont {G.}~\bibnamefont {Mazzola}},
  \bibinfo {author} {\bibfnamefont {J.}~\bibnamefont {Carrasquilla}}, \bibinfo
  {author} {\bibfnamefont {M.}~\bibnamefont {Troyer}}, \bibinfo {author}
  {\bibfnamefont {R.}~\bibnamefont {Melko}},\ and\ \bibinfo {author}
  {\bibfnamefont {G.}~\bibnamefont {Carleo}},\ }\href
  {https://doi.org/10.1038/s41567-018-0048-5} {\bibfield  {journal} {\bibinfo
  {journal} {Nature Physics}\ }\textbf {\bibinfo {volume} {14}},\ \bibinfo
  {pages} {447} (\bibinfo {year} {2018})}\BibitemShut {NoStop}%
\bibitem [{\citenamefont {Hsieh}\ \emph {et~al.}(2022)\citenamefont {Hsieh},
  \citenamefont {Chen}, \citenamefont {Wu}, \citenamefont {Chen}, \citenamefont
  {Ning}, \citenamefont {Huang}, \citenamefont {Wu},\ and\ \citenamefont
  {Lee}}]{Hsieh22}%
  \BibitemOpen
  \bibfield  {author} {\bibinfo {author} {\bibfnamefont {H.-Y.}\ \bibnamefont
  {Hsieh}}, \bibinfo {author} {\bibfnamefont {Y.-R.}\ \bibnamefont {Chen}},
  \bibinfo {author} {\bibfnamefont {H.-C.}\ \bibnamefont {Wu}}, \bibinfo
  {author} {\bibfnamefont {H.~L.}\ \bibnamefont {Chen}}, \bibinfo {author}
  {\bibfnamefont {J.}~\bibnamefont {Ning}}, \bibinfo {author} {\bibfnamefont
  {Y.-C.}\ \bibnamefont {Huang}}, \bibinfo {author} {\bibfnamefont {C.-M.}\
  \bibnamefont {Wu}},\ and\ \bibinfo {author} {\bibfnamefont {R.-K.}\
  \bibnamefont {Lee}},\ }\href {https://doi.org/10.1103/PhysRevLett.128.073604}
  {\bibfield  {journal} {\bibinfo  {journal} {Phys. Rev. Lett.}\ }\textbf
  {\bibinfo {volume} {128}},\ \bibinfo {pages} {073604} (\bibinfo {year}
  {2022})}\BibitemShut {NoStop}%
\bibitem [{\citenamefont {Vendeiro}\ \emph {et~al.}(2022)\citenamefont
  {Vendeiro}, \citenamefont {Ramette}, \citenamefont {Rudelis}, \citenamefont
  {Chong}, \citenamefont {Sinclair}, \citenamefont {Stewart}, \citenamefont
  {Urvoy},\ and\ \citenamefont {Vuletić}}]{Vendeiro2022}%
  \BibitemOpen
  \bibfield  {author} {\bibinfo {author} {\bibfnamefont {Z.}~\bibnamefont
  {Vendeiro}}, \bibinfo {author} {\bibfnamefont {J.}~\bibnamefont {Ramette}},
  \bibinfo {author} {\bibfnamefont {A.}~\bibnamefont {Rudelis}}, \bibinfo
  {author} {\bibfnamefont {M.}~\bibnamefont {Chong}}, \bibinfo {author}
  {\bibfnamefont {J.}~\bibnamefont {Sinclair}}, \bibinfo {author}
  {\bibfnamefont {L.}~\bibnamefont {Stewart}}, \bibinfo {author} {\bibfnamefont
  {A.}~\bibnamefont {Urvoy}},\ and\ \bibinfo {author} {\bibfnamefont
  {V.}~\bibnamefont {Vuletić}},\ }\href
  {https://doi.org/10.48550/ARXIV.2205.08057} {\bibinfo {title}
  {Machine-learning-accelerated bose-einstein condensation}} (\bibinfo {year}
  {2022})\BibitemShut {NoStop}%
\bibitem [{\citenamefont {Hentschel}\ and\ \citenamefont
  {Sanders}(2010)}]{Hentschel10}%
  \BibitemOpen
  \bibfield  {author} {\bibinfo {author} {\bibfnamefont {A.}~\bibnamefont
  {Hentschel}}\ and\ \bibinfo {author} {\bibfnamefont {B.~C.}\ \bibnamefont
  {Sanders}},\ }\href {https://doi.org/10.1103/PhysRevLett.104.063603}
  {\bibfield  {journal} {\bibinfo  {journal} {Phys. Rev. Lett.}\ }\textbf
  {\bibinfo {volume} {104}},\ \bibinfo {pages} {063603} (\bibinfo {year}
  {2010})}\BibitemShut {NoStop}%
\bibitem [{\citenamefont {Collaboration}(2015)}]{aLIGOdesign}%
  \BibitemOpen
  \bibfield  {author} {\bibinfo {author} {\bibfnamefont {T.~L.~S.}\
  \bibnamefont {Collaboration}},\ }\href
  {https://doi.org/10.1088/0264-9381/32/7/074001} {\bibfield  {journal}
  {\bibinfo  {journal} {Classical and Quantum Gravity}\ }\textbf {\bibinfo
  {volume} {32}},\ \bibinfo {pages} {074001} (\bibinfo {year}
  {2015})}\BibitemShut {NoStop}%
\bibitem [{\citenamefont {Acernese}\ \emph {et~al.}(2014)\citenamefont
  {Acernese} \emph {et~al.}}]{VirgoDesign}%
  \BibitemOpen
  \bibfield  {author} {\bibinfo {author} {\bibfnamefont {F.}~\bibnamefont
  {Acernese}} \emph {et~al.},\ }\href
  {https://doi.org/10.1088/0264-9381/32/2/024001} {\bibfield  {journal}
  {\bibinfo  {journal} {Class. Quantum Grav.}\ }\textbf {\bibinfo {volume}
  {32}},\ \bibinfo {pages} {024001} (\bibinfo {year} {2014})}\BibitemShut
  {NoStop}%
\bibitem [{\citenamefont {Dooley}\ \emph {et~al.}(2016)\citenamefont {Dooley},
  \citenamefont {Leong}, \citenamefont {Adams}, \citenamefont {Affeldt},
  \citenamefont {Bisht}, \citenamefont {Bogan}, \citenamefont {Degallaix},
  \citenamefont {Gr\"{a}f}, \citenamefont {Hild}, \citenamefont {Hough},
  \citenamefont {Khalaidovski}, \citenamefont {Lastzka}, \citenamefont {Lough},
  \citenamefont {L\"{u}ck}, \citenamefont {Macleod}, \citenamefont {Nuttall},
  \citenamefont {Prijatelj}, \citenamefont {Schnabel}, \citenamefont
  {Schreiber}, \citenamefont {Slutsky}, \citenamefont {Sorazu}, \citenamefont
  {Strain}, \citenamefont {Vahlbruch}, \citenamefont {W\k{a}s}, \citenamefont
  {Willke}, \citenamefont {Wittel}, \citenamefont {Danzmann},\ and\
  \citenamefont {Grote}}]{Dooley_2016}%
  \BibitemOpen
  \bibfield  {author} {\bibinfo {author} {\bibfnamefont {K.~L.}\ \bibnamefont
  {Dooley}}, \bibinfo {author} {\bibfnamefont {J.~R.}\ \bibnamefont {Leong}},
  \bibinfo {author} {\bibfnamefont {T.}~\bibnamefont {Adams}}, \bibinfo
  {author} {\bibfnamefont {C.}~\bibnamefont {Affeldt}}, \bibinfo {author}
  {\bibfnamefont {A.}~\bibnamefont {Bisht}}, \bibinfo {author} {\bibfnamefont
  {C.}~\bibnamefont {Bogan}}, \bibinfo {author} {\bibfnamefont
  {J.}~\bibnamefont {Degallaix}}, \bibinfo {author} {\bibfnamefont
  {C.}~\bibnamefont {Gr\"{a}f}}, \bibinfo {author} {\bibfnamefont
  {S.}~\bibnamefont {Hild}}, \bibinfo {author} {\bibfnamefont {J.}~\bibnamefont
  {Hough}}, \bibinfo {author} {\bibfnamefont {A.}~\bibnamefont {Khalaidovski}},
  \bibinfo {author} {\bibfnamefont {N.}~\bibnamefont {Lastzka}}, \bibinfo
  {author} {\bibfnamefont {J.}~\bibnamefont {Lough}}, \bibinfo {author}
  {\bibfnamefont {H.}~\bibnamefont {L\"{u}ck}}, \bibinfo {author}
  {\bibfnamefont {D.}~\bibnamefont {Macleod}}, \bibinfo {author} {\bibfnamefont
  {L.}~\bibnamefont {Nuttall}}, \bibinfo {author} {\bibfnamefont
  {M.}~\bibnamefont {Prijatelj}}, \bibinfo {author} {\bibfnamefont
  {R.}~\bibnamefont {Schnabel}}, \bibinfo {author} {\bibfnamefont
  {E.}~\bibnamefont {Schreiber}}, \bibinfo {author} {\bibfnamefont
  {J.}~\bibnamefont {Slutsky}}, \bibinfo {author} {\bibfnamefont
  {B.}~\bibnamefont {Sorazu}}, \bibinfo {author} {\bibfnamefont {K.~A.}\
  \bibnamefont {Strain}}, \bibinfo {author} {\bibfnamefont {H.}~\bibnamefont
  {Vahlbruch}}, \bibinfo {author} {\bibfnamefont {M.}~\bibnamefont {W\k{a}s}},
  \bibinfo {author} {\bibfnamefont {B.}~\bibnamefont {Willke}}, \bibinfo
  {author} {\bibfnamefont {H.}~\bibnamefont {Wittel}}, \bibinfo {author}
  {\bibfnamefont {K.}~\bibnamefont {Danzmann}},\ and\ \bibinfo {author}
  {\bibfnamefont {H.}~\bibnamefont {Grote}},\ }\href
  {https://doi.org/10.1088/0264-9381/33/7/075009} {\bibfield  {journal}
  {\bibinfo  {journal} {Classical and Quantum Gravity}\ }\textbf {\bibinfo
  {volume} {33}},\ \bibinfo {pages} {075009} (\bibinfo {year}
  {2016})}\BibitemShut {NoStop}%
\bibitem [{\citenamefont {Akutsu}\ \emph {et~al.}(2019)\citenamefont {Akutsu},
  \citenamefont {Ando}, \citenamefont {Arai}, \citenamefont {Arai},\ and\
  \citenamefont {collaboration}}]{Akutsu2019}%
  \BibitemOpen
  \bibfield  {author} {\bibinfo {author} {\bibfnamefont {T.}~\bibnamefont
  {Akutsu}}, \bibinfo {author} {\bibfnamefont {M.}~\bibnamefont {Ando}},
  \bibinfo {author} {\bibfnamefont {K.}~\bibnamefont {Arai}}, \bibinfo {author}
  {\bibfnamefont {Y.}~\bibnamefont {Arai}},\ and\ \bibinfo {author}
  {\bibfnamefont {K.}~\bibnamefont {collaboration}},\ }\href
  {https://doi.org/10.1038/s41550-018-0658-y} {\bibfield  {journal} {\bibinfo
  {journal} {Nature Astronomy}\ }\textbf {\bibinfo {volume} {3}},\ \bibinfo
  {pages} {35} (\bibinfo {year} {2019})}\BibitemShut {NoStop}%
\bibitem [{\citenamefont {Biswas}\ \emph {et~al.}(2013)\citenamefont {Biswas},
  \citenamefont {Blackburn}, \citenamefont {Cao}, \citenamefont {Essick},
  \citenamefont {Hodge}, \citenamefont {Katsavounidis}, \citenamefont {Kim},
  \citenamefont {Kim}, \citenamefont {Le~Bigot}, \citenamefont {Lee},
  \citenamefont {Oh}, \citenamefont {Oh}, \citenamefont {Son}, \citenamefont
  {Tao}, \citenamefont {Vaulin},\ and\ \citenamefont
  {Wang}}]{PhysRevD.88.062003}%
  \BibitemOpen
  \bibfield  {author} {\bibinfo {author} {\bibfnamefont {R.}~\bibnamefont
  {Biswas}}, \bibinfo {author} {\bibfnamefont {L.}~\bibnamefont {Blackburn}},
  \bibinfo {author} {\bibfnamefont {J.}~\bibnamefont {Cao}}, \bibinfo {author}
  {\bibfnamefont {R.}~\bibnamefont {Essick}}, \bibinfo {author} {\bibfnamefont
  {K.~A.}\ \bibnamefont {Hodge}}, \bibinfo {author} {\bibfnamefont
  {E.}~\bibnamefont {Katsavounidis}}, \bibinfo {author} {\bibfnamefont
  {K.}~\bibnamefont {Kim}}, \bibinfo {author} {\bibfnamefont {Y.-M.}\
  \bibnamefont {Kim}}, \bibinfo {author} {\bibfnamefont {E.-O.}\ \bibnamefont
  {Le~Bigot}}, \bibinfo {author} {\bibfnamefont {C.-H.}\ \bibnamefont {Lee}},
  \bibinfo {author} {\bibfnamefont {J.~J.}\ \bibnamefont {Oh}}, \bibinfo
  {author} {\bibfnamefont {S.~H.}\ \bibnamefont {Oh}}, \bibinfo {author}
  {\bibfnamefont {E.~J.}\ \bibnamefont {Son}}, \bibinfo {author} {\bibfnamefont
  {Y.}~\bibnamefont {Tao}}, \bibinfo {author} {\bibfnamefont {R.}~\bibnamefont
  {Vaulin}},\ and\ \bibinfo {author} {\bibfnamefont {X.}~\bibnamefont {Wang}},\
  }\href {https://doi.org/10.1103/PhysRevD.88.062003} {\bibfield  {journal}
  {\bibinfo  {journal} {Phys. Rev. D}\ }\textbf {\bibinfo {volume} {88}},\
  \bibinfo {pages} {062003} (\bibinfo {year} {2013})}\BibitemShut {NoStop}%
\bibitem [{\citenamefont {Powell}\ \emph {et~al.}(2015)\citenamefont {Powell},
  \citenamefont {Trifirò}, \citenamefont {Cuoco}, \citenamefont {Heng},\ and\
  \citenamefont {Cavaglià}}]{Powell_2015}%
  \BibitemOpen
  \bibfield  {author} {\bibinfo {author} {\bibfnamefont {J.}~\bibnamefont
  {Powell}}, \bibinfo {author} {\bibfnamefont {D.}~\bibnamefont {Trifirò}},
  \bibinfo {author} {\bibfnamefont {E.}~\bibnamefont {Cuoco}}, \bibinfo
  {author} {\bibfnamefont {I.~S.}\ \bibnamefont {Heng}},\ and\ \bibinfo
  {author} {\bibfnamefont {M.}~\bibnamefont {Cavaglià}},\ }\href
  {https://doi.org/10.1088/0264-9381/32/21/215012} {\bibfield  {journal}
  {\bibinfo  {journal} {Classical and Quantum Gravity}\ }\textbf {\bibinfo
  {volume} {32}},\ \bibinfo {pages} {215012} (\bibinfo {year}
  {2015})}\BibitemShut {NoStop}%
\bibitem [{\citenamefont {Zevin}\ \emph {et~al.}(2017)\citenamefont {Zevin},
  \citenamefont {Coughlin}, \citenamefont {Bahaadini}, \citenamefont {Besler},
  \citenamefont {Rohani}, \citenamefont {Allen}, \citenamefont {Cabero},
  \citenamefont {Crowston}, \citenamefont {Katsaggelos}, \citenamefont
  {Larson}, \citenamefont {Lee}, \citenamefont {Lintott}, \citenamefont
  {Littenberg}, \citenamefont {Lundgren}, \citenamefont {Østerlund},
  \citenamefont {Smith}, \citenamefont {Trouille},\ and\ \citenamefont
  {Kalogera}}]{Zevin_2017}%
  \BibitemOpen
  \bibfield  {author} {\bibinfo {author} {\bibfnamefont {M.}~\bibnamefont
  {Zevin}}, \bibinfo {author} {\bibfnamefont {S.}~\bibnamefont {Coughlin}},
  \bibinfo {author} {\bibfnamefont {S.}~\bibnamefont {Bahaadini}}, \bibinfo
  {author} {\bibfnamefont {E.}~\bibnamefont {Besler}}, \bibinfo {author}
  {\bibfnamefont {N.}~\bibnamefont {Rohani}}, \bibinfo {author} {\bibfnamefont
  {S.}~\bibnamefont {Allen}}, \bibinfo {author} {\bibfnamefont
  {M.}~\bibnamefont {Cabero}}, \bibinfo {author} {\bibfnamefont
  {K.}~\bibnamefont {Crowston}}, \bibinfo {author} {\bibfnamefont {A.~K.}\
  \bibnamefont {Katsaggelos}}, \bibinfo {author} {\bibfnamefont {S.~L.}\
  \bibnamefont {Larson}}, \bibinfo {author} {\bibfnamefont {T.~K.}\
  \bibnamefont {Lee}}, \bibinfo {author} {\bibfnamefont {C.}~\bibnamefont
  {Lintott}}, \bibinfo {author} {\bibfnamefont {T.~B.}\ \bibnamefont
  {Littenberg}}, \bibinfo {author} {\bibfnamefont {A.}~\bibnamefont
  {Lundgren}}, \bibinfo {author} {\bibfnamefont {C.}~\bibnamefont
  {Østerlund}}, \bibinfo {author} {\bibfnamefont {J.~R.}\ \bibnamefont
  {Smith}}, \bibinfo {author} {\bibfnamefont {L.}~\bibnamefont {Trouille}},\
  and\ \bibinfo {author} {\bibfnamefont {V.}~\bibnamefont {Kalogera}},\ }\href
  {https://doi.org/10.1088/1361-6382/aa5cea} {\bibfield  {journal} {\bibinfo
  {journal} {Classical and Quantum Gravity}\ }\textbf {\bibinfo {volume}
  {34}},\ \bibinfo {pages} {064003} (\bibinfo {year} {2017})}\BibitemShut
  {NoStop}%
\bibitem [{\citenamefont {Mukund}\ \emph {et~al.}(2017)\citenamefont {Mukund},
  \citenamefont {Abraham}, \citenamefont {Kandhasamy}, \citenamefont {Mitra},\
  and\ \citenamefont {Philip}}]{PhysRevD.95.104059}%
  \BibitemOpen
  \bibfield  {author} {\bibinfo {author} {\bibfnamefont {N.}~\bibnamefont
  {Mukund}}, \bibinfo {author} {\bibfnamefont {S.}~\bibnamefont {Abraham}},
  \bibinfo {author} {\bibfnamefont {S.}~\bibnamefont {Kandhasamy}}, \bibinfo
  {author} {\bibfnamefont {S.}~\bibnamefont {Mitra}},\ and\ \bibinfo {author}
  {\bibfnamefont {N.~S.}\ \bibnamefont {Philip}},\ }\href
  {https://doi.org/10.1103/PhysRevD.95.104059} {\bibfield  {journal} {\bibinfo
  {journal} {Phys. Rev. D}\ }\textbf {\bibinfo {volume} {95}},\ \bibinfo
  {pages} {104059} (\bibinfo {year} {2017})}\BibitemShut {NoStop}%
\bibitem [{\citenamefont {Ormiston}\ \emph {et~al.}(2020)\citenamefont
  {Ormiston}, \citenamefont {Nguyen}, \citenamefont {Coughlin}, \citenamefont
  {Adhikari},\ and\ \citenamefont {Katsavounidis}}]{Ormiston2020}%
  \BibitemOpen
  \bibfield  {author} {\bibinfo {author} {\bibfnamefont {R.}~\bibnamefont
  {Ormiston}}, \bibinfo {author} {\bibfnamefont {T.}~\bibnamefont {Nguyen}},
  \bibinfo {author} {\bibfnamefont {M.}~\bibnamefont {Coughlin}}, \bibinfo
  {author} {\bibfnamefont {R.~X.}\ \bibnamefont {Adhikari}},\ and\ \bibinfo
  {author} {\bibfnamefont {E.}~\bibnamefont {Katsavounidis}},\ }\href
  {https://doi.org/10.1103/PhysRevResearch.2.033066} {\bibfield  {journal}
  {\bibinfo  {journal} {Phys. Rev. Research}\ }\textbf {\bibinfo {volume}
  {2}},\ \bibinfo {pages} {033066} (\bibinfo {year} {2020})}\BibitemShut
  {NoStop}%
\bibitem [{\citenamefont {Vajente}\ \emph {et~al.}(2020)\citenamefont
  {Vajente}, \citenamefont {Huang}, \citenamefont {Isi}, \citenamefont
  {Driggers}, \citenamefont {Kissel}, \citenamefont
  {Szczepa\ifmmode~\acute{n}\else \'{n}\fi{}czyk},\ and\ \citenamefont
  {Vitale}}]{Vajente2020}%
  \BibitemOpen
  \bibfield  {author} {\bibinfo {author} {\bibfnamefont {G.}~\bibnamefont
  {Vajente}}, \bibinfo {author} {\bibfnamefont {Y.}~\bibnamefont {Huang}},
  \bibinfo {author} {\bibfnamefont {M.}~\bibnamefont {Isi}}, \bibinfo {author}
  {\bibfnamefont {J.~C.}\ \bibnamefont {Driggers}}, \bibinfo {author}
  {\bibfnamefont {J.~S.}\ \bibnamefont {Kissel}}, \bibinfo {author}
  {\bibfnamefont {M.~J.}\ \bibnamefont {Szczepa\ifmmode~\acute{n}\else
  \'{n}\fi{}czyk}},\ and\ \bibinfo {author} {\bibfnamefont {S.}~\bibnamefont
  {Vitale}},\ }\href {https://doi.org/10.1103/PhysRevD.101.042003} {\bibfield
  {journal} {\bibinfo  {journal} {Phys. Rev. D}\ }\textbf {\bibinfo {volume}
  {101}},\ \bibinfo {pages} {042003} (\bibinfo {year} {2020})}\BibitemShut
  {NoStop}%
\bibitem [{\citenamefont {Yu}\ and\ \citenamefont {Adhikari}(2022)}]{Hang2022}%
  \BibitemOpen
  \bibfield  {author} {\bibinfo {author} {\bibfnamefont {H.}~\bibnamefont
  {Yu}}\ and\ \bibinfo {author} {\bibfnamefont {R.~X.}\ \bibnamefont
  {Adhikari}},\ }\href {https://doi.org/10.3389/frai.2022.811563} {\bibfield
  {journal} {\bibinfo  {journal} {Front. Artif. Intell.}\ }\textbf {\bibinfo
  {volume} {5}},\ \bibinfo {pages} {811563} (\bibinfo {year}
  {2022})}\BibitemShut {NoStop}%
\bibitem [{\citenamefont {Doctor}\ \emph {et~al.}(2017)\citenamefont {Doctor},
  \citenamefont {Farr}, \citenamefont {Holz},\ and\ \citenamefont
  {P\"urrer}}]{Doctor2017}%
  \BibitemOpen
  \bibfield  {author} {\bibinfo {author} {\bibfnamefont {Z.}~\bibnamefont
  {Doctor}}, \bibinfo {author} {\bibfnamefont {B.}~\bibnamefont {Farr}},
  \bibinfo {author} {\bibfnamefont {D.~E.}\ \bibnamefont {Holz}},\ and\
  \bibinfo {author} {\bibfnamefont {M.}~\bibnamefont {P\"urrer}},\ }\href
  {https://doi.org/10.1103/PhysRevD.96.123011} {\bibfield  {journal} {\bibinfo
  {journal} {Phys. Rev. D}\ }\textbf {\bibinfo {volume} {96}},\ \bibinfo
  {pages} {123011} (\bibinfo {year} {2017})}\BibitemShut {NoStop}%
\bibitem [{\citenamefont {Baker}\ \emph {et~al.}(2015)\citenamefont {Baker},
  \citenamefont {Caudill}, \citenamefont {Hodge}, \citenamefont {Talukder},
  \citenamefont {Capano},\ and\ \citenamefont {Cornish}}]{Baker2015}%
  \BibitemOpen
  \bibfield  {author} {\bibinfo {author} {\bibfnamefont {P.~T.}\ \bibnamefont
  {Baker}}, \bibinfo {author} {\bibfnamefont {S.}~\bibnamefont {Caudill}},
  \bibinfo {author} {\bibfnamefont {K.~A.}\ \bibnamefont {Hodge}}, \bibinfo
  {author} {\bibfnamefont {D.}~\bibnamefont {Talukder}}, \bibinfo {author}
  {\bibfnamefont {C.}~\bibnamefont {Capano}},\ and\ \bibinfo {author}
  {\bibfnamefont {N.~J.}\ \bibnamefont {Cornish}},\ }\href
  {https://doi.org/10.1103/PhysRevD.91.062004} {\bibfield  {journal} {\bibinfo
  {journal} {Phys. Rev. D}\ }\textbf {\bibinfo {volume} {91}},\ \bibinfo
  {pages} {062004} (\bibinfo {year} {2015})}\BibitemShut {NoStop}%
\bibitem [{\citenamefont {Mukund}\ \emph {et~al.}(2023)\citenamefont {Mukund},
  \citenamefont {Lough}, \citenamefont {Bisht}, \citenamefont {Wittel},
  \citenamefont {Nadji}, \citenamefont {Affeldt}, \citenamefont {Bergamin},
  \citenamefont {Brinkmann}, \citenamefont {Kringel}, \citenamefont {Lück},
  \citenamefont {Weinert},\ and\ \citenamefont
  {Danzmann}}]{mukund2023demonstration}%
  \BibitemOpen
  \bibfield  {author} {\bibinfo {author} {\bibfnamefont {N.}~\bibnamefont
  {Mukund}}, \bibinfo {author} {\bibfnamefont {J.}~\bibnamefont {Lough}},
  \bibinfo {author} {\bibfnamefont {A.}~\bibnamefont {Bisht}}, \bibinfo
  {author} {\bibfnamefont {H.}~\bibnamefont {Wittel}}, \bibinfo {author}
  {\bibfnamefont {S.~L.}\ \bibnamefont {Nadji}}, \bibinfo {author}
  {\bibfnamefont {C.}~\bibnamefont {Affeldt}}, \bibinfo {author} {\bibfnamefont
  {F.}~\bibnamefont {Bergamin}}, \bibinfo {author} {\bibfnamefont
  {M.}~\bibnamefont {Brinkmann}}, \bibinfo {author} {\bibfnamefont
  {V.}~\bibnamefont {Kringel}}, \bibinfo {author} {\bibfnamefont
  {H.}~\bibnamefont {Lück}}, \bibinfo {author} {\bibfnamefont
  {M.}~\bibnamefont {Weinert}},\ and\ \bibinfo {author} {\bibfnamefont
  {K.}~\bibnamefont {Danzmann}},\ }\href@noop {} {\bibinfo {title} {First
  demonstration of neural sensing and control in a kilometer-scale
  gravitational wave observatory}} (\bibinfo {year} {2023}),\ \Eprint
  {https://arxiv.org/abs/2301.06221} {arXiv:2301.06221 [physics.ins-det]}
  \BibitemShut {NoStop}%
\bibitem [{\citenamefont {{M. Tse, {\it et al.}}}(2019)}]{Tse2019}%
  \BibitemOpen
  \bibfield  {author} {\bibinfo {author} {\bibnamefont {{M. Tse, {\it et
  al.}}}},\ }\href {https://doi.org/10.1103/PhysRevLett.123.231107} {\bibfield
  {journal} {\bibinfo  {journal} {Phys. Rev. Lett.}\ }\textbf {\bibinfo
  {volume} {123}},\ \bibinfo {pages} {231107} (\bibinfo {year}
  {2019})}\BibitemShut {NoStop}%
\bibitem [{\citenamefont {{F. Acernese, {\it et al.}}}(2019)}]{Acernese2019}%
  \BibitemOpen
  \bibfield  {author} {\bibinfo {author} {\bibnamefont {{F. Acernese, {\it et
  al.}}}} (\bibinfo {collaboration} {Virgo Collaboration}),\ }\href
  {https://doi.org/10.1103/PhysRevLett.123.231108} {\bibfield  {journal}
  {\bibinfo  {journal} {Phys. Rev. Lett.}\ }\textbf {\bibinfo {volume} {123}},\
  \bibinfo {pages} {231108} (\bibinfo {year} {2019})}\BibitemShut {NoStop}%
\bibitem [{\citenamefont {Lough}\ \emph {et~al.}(2021)\citenamefont {Lough},
  \citenamefont {Schreiber}, \citenamefont {Bergamin}, \citenamefont {Grote},
  \citenamefont {Mehmet}, \citenamefont {Vahlbruch}, \citenamefont {Affeldt},
  \citenamefont {Brinkmann}, \citenamefont {Bisht}, \citenamefont {Kringel}
  \emph {et~al.}}]{Lough2021}%
  \BibitemOpen
  \bibfield  {author} {\bibinfo {author} {\bibfnamefont {J.}~\bibnamefont
  {Lough}}, \bibinfo {author} {\bibfnamefont {E.}~\bibnamefont {Schreiber}},
  \bibinfo {author} {\bibfnamefont {F.}~\bibnamefont {Bergamin}}, \bibinfo
  {author} {\bibfnamefont {H.}~\bibnamefont {Grote}}, \bibinfo {author}
  {\bibfnamefont {M.}~\bibnamefont {Mehmet}}, \bibinfo {author} {\bibfnamefont
  {H.}~\bibnamefont {Vahlbruch}}, \bibinfo {author} {\bibfnamefont
  {C.}~\bibnamefont {Affeldt}}, \bibinfo {author} {\bibfnamefont
  {M.}~\bibnamefont {Brinkmann}}, \bibinfo {author} {\bibfnamefont
  {A.}~\bibnamefont {Bisht}}, \bibinfo {author} {\bibfnamefont
  {V.}~\bibnamefont {Kringel}}, \emph {et~al.},\ }\href
  {https://doi.org/10.1103/PhysRevLett.126.041102} {\bibfield  {journal}
  {\bibinfo  {journal} {Physical Review Letters}\ }\textbf {\bibinfo {volume}
  {126}},\ \bibinfo {pages} {041102} (\bibinfo {year} {2021})}\BibitemShut
  {NoStop}%
\bibitem [{\citenamefont {Barsotti}\ \emph {et~al.}(2018)\citenamefont
  {Barsotti}, \citenamefont {Harms},\ and\ \citenamefont
  {Schnabel}}]{Barsotti_2018}%
  \BibitemOpen
  \bibfield  {author} {\bibinfo {author} {\bibfnamefont {L.}~\bibnamefont
  {Barsotti}}, \bibinfo {author} {\bibfnamefont {J.}~\bibnamefont {Harms}},\
  and\ \bibinfo {author} {\bibfnamefont {R.}~\bibnamefont {Schnabel}},\ }\href
  {https://doi.org/10.1088/1361-6633/aab906} {\bibfield  {journal} {\bibinfo
  {journal} {Reports on Progress in Physics}\ }\textbf {\bibinfo {volume}
  {82}},\ \bibinfo {pages} {016905} (\bibinfo {year} {2018})}\BibitemShut
  {NoStop}%
\bibitem [{\citenamefont {McCuller}\ \emph {et~al.}(2020)\citenamefont
  {McCuller}, \citenamefont {Whittle}, \citenamefont {Ganapathy}, \citenamefont
  {Komori}, \citenamefont {Tse}, \citenamefont {Fernandez-Galiana},
  \citenamefont {Barsotti}, \citenamefont {Fritschel}, \citenamefont
  {MacInnis}, \citenamefont {Matichard}, \citenamefont {Mason}, \citenamefont
  {Mavalvala}, \citenamefont {Mittleman}, \citenamefont {Yu}, \citenamefont
  {Zucker},\ and\ \citenamefont {Evans}}]{McCuller2020}%
  \BibitemOpen
  \bibfield  {author} {\bibinfo {author} {\bibfnamefont {L.}~\bibnamefont
  {McCuller}}, \bibinfo {author} {\bibfnamefont {C.}~\bibnamefont {Whittle}},
  \bibinfo {author} {\bibfnamefont {D.}~\bibnamefont {Ganapathy}}, \bibinfo
  {author} {\bibfnamefont {K.}~\bibnamefont {Komori}}, \bibinfo {author}
  {\bibfnamefont {M.}~\bibnamefont {Tse}}, \bibinfo {author} {\bibfnamefont
  {A.}~\bibnamefont {Fernandez-Galiana}}, \bibinfo {author} {\bibfnamefont
  {L.}~\bibnamefont {Barsotti}}, \bibinfo {author} {\bibfnamefont
  {P.}~\bibnamefont {Fritschel}}, \bibinfo {author} {\bibfnamefont
  {M.}~\bibnamefont {MacInnis}}, \bibinfo {author} {\bibfnamefont
  {F.}~\bibnamefont {Matichard}}, \bibinfo {author} {\bibfnamefont
  {K.}~\bibnamefont {Mason}}, \bibinfo {author} {\bibfnamefont
  {N.}~\bibnamefont {Mavalvala}}, \bibinfo {author} {\bibfnamefont
  {R.}~\bibnamefont {Mittleman}}, \bibinfo {author} {\bibfnamefont
  {H.}~\bibnamefont {Yu}}, \bibinfo {author} {\bibfnamefont {M.~E.}\
  \bibnamefont {Zucker}},\ and\ \bibinfo {author} {\bibfnamefont
  {M.}~\bibnamefont {Evans}},\ }\href
  {https://doi.org/10.1103/PhysRevLett.124.171102} {\bibfield  {journal}
  {\bibinfo  {journal} {Phys. Rev. Lett.}\ }\textbf {\bibinfo {volume} {124}},\
  \bibinfo {pages} {171102} (\bibinfo {year} {2020})}\BibitemShut {NoStop}%
\bibitem [{\citenamefont {Zhao}\ \emph {et~al.}(2020)\citenamefont {Zhao},
  \citenamefont {Aritomi}, \citenamefont {Capocasa}, \citenamefont {Leonardi},
  \citenamefont {Eisenmann}, \citenamefont {Guo}, \citenamefont {Polini},
  \citenamefont {Tomura}, \citenamefont {Arai}, \citenamefont {Aso},
  \citenamefont {Huang}, \citenamefont {Lee}, \citenamefont {L\"uck},
  \citenamefont {Miyakawa}, \citenamefont {Prat}, \citenamefont {Shoda},
  \citenamefont {Tacca}, \citenamefont {Takahashi}, \citenamefont {Vahlbruch},
  \citenamefont {Vardaro}, \citenamefont {Wu}, \citenamefont {Barsuglia},\ and\
  \citenamefont {Flaminio}}]{Zhao2020}%
  \BibitemOpen
  \bibfield  {author} {\bibinfo {author} {\bibfnamefont {Y.}~\bibnamefont
  {Zhao}}, \bibinfo {author} {\bibfnamefont {N.}~\bibnamefont {Aritomi}},
  \bibinfo {author} {\bibfnamefont {E.}~\bibnamefont {Capocasa}}, \bibinfo
  {author} {\bibfnamefont {M.}~\bibnamefont {Leonardi}}, \bibinfo {author}
  {\bibfnamefont {M.}~\bibnamefont {Eisenmann}}, \bibinfo {author}
  {\bibfnamefont {Y.}~\bibnamefont {Guo}}, \bibinfo {author} {\bibfnamefont
  {E.}~\bibnamefont {Polini}}, \bibinfo {author} {\bibfnamefont
  {A.}~\bibnamefont {Tomura}}, \bibinfo {author} {\bibfnamefont
  {K.}~\bibnamefont {Arai}}, \bibinfo {author} {\bibfnamefont {Y.}~\bibnamefont
  {Aso}}, \bibinfo {author} {\bibfnamefont {Y.-C.}\ \bibnamefont {Huang}},
  \bibinfo {author} {\bibfnamefont {R.-K.}\ \bibnamefont {Lee}}, \bibinfo
  {author} {\bibfnamefont {H.}~\bibnamefont {L\"uck}}, \bibinfo {author}
  {\bibfnamefont {O.}~\bibnamefont {Miyakawa}}, \bibinfo {author}
  {\bibfnamefont {P.}~\bibnamefont {Prat}}, \bibinfo {author} {\bibfnamefont
  {A.}~\bibnamefont {Shoda}}, \bibinfo {author} {\bibfnamefont
  {M.}~\bibnamefont {Tacca}}, \bibinfo {author} {\bibfnamefont
  {R.}~\bibnamefont {Takahashi}}, \bibinfo {author} {\bibfnamefont
  {H.}~\bibnamefont {Vahlbruch}}, \bibinfo {author} {\bibfnamefont
  {M.}~\bibnamefont {Vardaro}}, \bibinfo {author} {\bibfnamefont {C.-M.}\
  \bibnamefont {Wu}}, \bibinfo {author} {\bibfnamefont {M.}~\bibnamefont
  {Barsuglia}},\ and\ \bibinfo {author} {\bibfnamefont {R.}~\bibnamefont
  {Flaminio}},\ }\href {https://doi.org/10.1103/PhysRevLett.124.171101}
  {\bibfield  {journal} {\bibinfo  {journal} {Phys. Rev. Lett.}\ }\textbf
  {\bibinfo {volume} {124}},\ \bibinfo {pages} {171101} (\bibinfo {year}
  {2020})}\BibitemShut {NoStop}%
\bibitem [{\citenamefont {Evans}\ \emph {et~al.}(2021)\citenamefont {Evans},
  \citenamefont {Adhikari}, \citenamefont {Afle}, \citenamefont {Ballmer},
  \citenamefont {Biscoveanu}, \citenamefont {Borhanian}, \citenamefont {Brown},
  \citenamefont {Chen}, \citenamefont {Eisenstein}, \citenamefont {Gruson},
  \citenamefont {Gupta}, \citenamefont {Hall}, \citenamefont {Huxford},
  \citenamefont {Kamai}, \citenamefont {Kashyap}, \citenamefont {Kissel},
  \citenamefont {Kuns}, \citenamefont {Landry}, \citenamefont {Lenon},
  \citenamefont {Lovelace}, \citenamefont {McCuller}, \citenamefont {Ng},
  \citenamefont {Nitz}, \citenamefont {Read}, \citenamefont {Sathyaprakash},
  \citenamefont {Shoemaker}, \citenamefont {Slagmolen}, \citenamefont {Smith},
  \citenamefont {Srivastava}, \citenamefont {Sun}, \citenamefont {Vitale},\
  and\ \citenamefont {Weiss}}]{CEHS}%
  \BibitemOpen
  \bibfield  {author} {\bibinfo {author} {\bibfnamefont {M.}~\bibnamefont
  {Evans}}, \bibinfo {author} {\bibfnamefont {R.~X.}\ \bibnamefont {Adhikari}},
  \bibinfo {author} {\bibfnamefont {C.}~\bibnamefont {Afle}}, \bibinfo {author}
  {\bibfnamefont {S.~W.}\ \bibnamefont {Ballmer}}, \bibinfo {author}
  {\bibfnamefont {S.}~\bibnamefont {Biscoveanu}}, \bibinfo {author}
  {\bibfnamefont {S.}~\bibnamefont {Borhanian}}, \bibinfo {author}
  {\bibfnamefont {D.~A.}\ \bibnamefont {Brown}}, \bibinfo {author}
  {\bibfnamefont {Y.}~\bibnamefont {Chen}}, \bibinfo {author} {\bibfnamefont
  {R.}~\bibnamefont {Eisenstein}}, \bibinfo {author} {\bibfnamefont
  {A.}~\bibnamefont {Gruson}}, \bibinfo {author} {\bibfnamefont
  {A.}~\bibnamefont {Gupta}}, \bibinfo {author} {\bibfnamefont {E.~D.}\
  \bibnamefont {Hall}}, \bibinfo {author} {\bibfnamefont {R.}~\bibnamefont
  {Huxford}}, \bibinfo {author} {\bibfnamefont {B.}~\bibnamefont {Kamai}},
  \bibinfo {author} {\bibfnamefont {R.}~\bibnamefont {Kashyap}}, \bibinfo
  {author} {\bibfnamefont {J.~S.}\ \bibnamefont {Kissel}}, \bibinfo {author}
  {\bibfnamefont {K.}~\bibnamefont {Kuns}}, \bibinfo {author} {\bibfnamefont
  {P.}~\bibnamefont {Landry}}, \bibinfo {author} {\bibfnamefont
  {A.}~\bibnamefont {Lenon}}, \bibinfo {author} {\bibfnamefont
  {G.}~\bibnamefont {Lovelace}}, \bibinfo {author} {\bibfnamefont
  {L.}~\bibnamefont {McCuller}}, \bibinfo {author} {\bibfnamefont {K.~K.~Y.}\
  \bibnamefont {Ng}}, \bibinfo {author} {\bibfnamefont {A.~H.}\ \bibnamefont
  {Nitz}}, \bibinfo {author} {\bibfnamefont {J.}~\bibnamefont {Read}}, \bibinfo
  {author} {\bibfnamefont {B.~S.}\ \bibnamefont {Sathyaprakash}}, \bibinfo
  {author} {\bibfnamefont {D.~H.}\ \bibnamefont {Shoemaker}}, \bibinfo {author}
  {\bibfnamefont {B.~J.~J.}\ \bibnamefont {Slagmolen}}, \bibinfo {author}
  {\bibfnamefont {J.~R.}\ \bibnamefont {Smith}}, \bibinfo {author}
  {\bibfnamefont {V.}~\bibnamefont {Srivastava}}, \bibinfo {author}
  {\bibfnamefont {L.}~\bibnamefont {Sun}}, \bibinfo {author} {\bibfnamefont
  {S.}~\bibnamefont {Vitale}},\ and\ \bibinfo {author} {\bibfnamefont
  {R.}~\bibnamefont {Weiss}},\ }\href@noop {} {\bibinfo {title} {A horizon
  study for {Cosmic Explorer}: Science, observatories, and community}}
  (\bibinfo {year} {2021}),\ \Eprint {https://arxiv.org/abs/2109.09882}
  {arXiv:2109.09882 [astro-ph.IM]} \BibitemShut {NoStop}%
\bibitem [{\citenamefont {Jones}\ \emph {et~al.}(2020)\citenamefont {Jones},
  \citenamefont {Zhang}, \citenamefont {Miao},\ and\ \citenamefont
  {Freise}}]{ET2020}%
  \BibitemOpen
  \bibfield  {author} {\bibinfo {author} {\bibfnamefont {P.}~\bibnamefont
  {Jones}}, \bibinfo {author} {\bibfnamefont {T.}~\bibnamefont {Zhang}},
  \bibinfo {author} {\bibfnamefont {H.}~\bibnamefont {Miao}},\ and\ \bibinfo
  {author} {\bibfnamefont {A.}~\bibnamefont {Freise}},\ }\href
  {https://doi.org/10.1103/PhysRevD.101.082002} {\bibfield  {journal} {\bibinfo
   {journal} {Phys. Rev. D}\ }\textbf {\bibinfo {volume} {101}},\ \bibinfo
  {pages} {082002} (\bibinfo {year} {2020})}\BibitemShut {NoStop}%
\bibitem [{\citenamefont {Slagmolen}\ \emph {et~al.}(2011)\citenamefont
  {Slagmolen}, \citenamefont {Mullavey}, \citenamefont {Miller}, \citenamefont
  {McClelland},\ and\ \citenamefont {Fritschel}}]{Slagmolen2011.RSI}%
  \BibitemOpen
  \bibfield  {author} {\bibinfo {author} {\bibfnamefont {B.~J.~J.}\
  \bibnamefont {Slagmolen}}, \bibinfo {author} {\bibfnamefont {A.~J.}\
  \bibnamefont {Mullavey}}, \bibinfo {author} {\bibfnamefont {J.}~\bibnamefont
  {Miller}}, \bibinfo {author} {\bibfnamefont {D.~E.}\ \bibnamefont
  {McClelland}},\ and\ \bibinfo {author} {\bibfnamefont {P.}~\bibnamefont
  {Fritschel}},\ }\href {https://doi.org/10.1063/1.3669532} {\bibfield
  {journal} {\bibinfo  {journal} {Review of Scientific Instruments}\ }\textbf
  {\bibinfo {volume} {82}},\ \bibinfo {eid} {125108} (\bibinfo {year}
  {2011})}\BibitemShut {NoStop}%
\bibitem [{\citenamefont {Vahlbruch}\ \emph {et~al.}(2006)\citenamefont
  {Vahlbruch}, \citenamefont {Chelkowski}, \citenamefont {Hage}, \citenamefont
  {Franzen}, \citenamefont {Danzmann},\ and\ \citenamefont
  {Schnabel}}]{Vahlbruch2006a}%
  \BibitemOpen
  \bibfield  {author} {\bibinfo {author} {\bibfnamefont {H.}~\bibnamefont
  {Vahlbruch}}, \bibinfo {author} {\bibfnamefont {S.}~\bibnamefont
  {Chelkowski}}, \bibinfo {author} {\bibfnamefont {B.}~\bibnamefont {Hage}},
  \bibinfo {author} {\bibfnamefont {A.}~\bibnamefont {Franzen}}, \bibinfo
  {author} {\bibfnamefont {K.}~\bibnamefont {Danzmann}},\ and\ \bibinfo
  {author} {\bibfnamefont {R.}~\bibnamefont {Schnabel}},\ }\href
  {https://doi.org/10.1103/PhysRevLett.97.011101} {\bibfield  {journal}
  {\bibinfo  {journal} {Physical Review Letters}\ }\textbf {\bibinfo {volume}
  {97}},\ \bibinfo {pages} {011101} (\bibinfo {year} {2006})}\BibitemShut
  {NoStop}%
\bibitem [{\citenamefont {Chelkowski}\ \emph {et~al.}(2007)\citenamefont
  {Chelkowski}, \citenamefont {Vahlbruch}, \citenamefont {Danzmann},\ and\
  \citenamefont {Schnabel}}]{Chelkowski2007}%
  \BibitemOpen
  \bibfield  {author} {\bibinfo {author} {\bibfnamefont {S.}~\bibnamefont
  {Chelkowski}}, \bibinfo {author} {\bibfnamefont {H.}~\bibnamefont
  {Vahlbruch}}, \bibinfo {author} {\bibfnamefont {K.}~\bibnamefont
  {Danzmann}},\ and\ \bibinfo {author} {\bibfnamefont {R.}~\bibnamefont
  {Schnabel}},\ }\href {https://doi.org/10.1103/PhysRevA.75.043814} {\bibfield
  {journal} {\bibinfo  {journal} {Physical Review A}\ }\textbf {\bibinfo
  {volume} {75}},\ \bibinfo {pages} {043814} (\bibinfo {year}
  {2007})}\BibitemShut {NoStop}%
\bibitem [{\citenamefont {Grote}\ \emph {et~al.}(2013)\citenamefont {Grote},
  \citenamefont {Danzmann}, \citenamefont {Dooley}, \citenamefont {Schnabel},
  \citenamefont {Slutsky},\ and\ \citenamefont {Vahlbruch}}]{GEOlongterm}%
  \BibitemOpen
  \bibfield  {author} {\bibinfo {author} {\bibfnamefont {H.}~\bibnamefont
  {Grote}}, \bibinfo {author} {\bibfnamefont {K.}~\bibnamefont {Danzmann}},
  \bibinfo {author} {\bibfnamefont {K.~L.}\ \bibnamefont {Dooley}}, \bibinfo
  {author} {\bibfnamefont {R.}~\bibnamefont {Schnabel}}, \bibinfo {author}
  {\bibfnamefont {J.}~\bibnamefont {Slutsky}},\ and\ \bibinfo {author}
  {\bibfnamefont {H.}~\bibnamefont {Vahlbruch}},\ }\href
  {https://doi.org/10.1103/PhysRevLett.110.181101} {\bibfield  {journal}
  {\bibinfo  {journal} {Phys. Rev. Lett.}\ }\textbf {\bibinfo {volume} {110}},\
  \bibinfo {pages} {181101} (\bibinfo {year} {2013})}\BibitemShut {NoStop}%
\bibitem [{\citenamefont {Schreiber}\ \emph {et~al.}(2016)\citenamefont
  {Schreiber}, \citenamefont {Dooley}, \citenamefont {Vahlbruch}, \citenamefont
  {Affeldt}, \citenamefont {Bisht}, \citenamefont {Leong}, \citenamefont
  {Lough}, \citenamefont {Prijatelj}, \citenamefont {Slutsky}, \citenamefont
  {Was}, \citenamefont {Wittel}, \citenamefont {Danzmann},\ and\ \citenamefont
  {Grote}}]{Schreiber:16}%
  \BibitemOpen
  \bibfield  {author} {\bibinfo {author} {\bibfnamefont {E.}~\bibnamefont
  {Schreiber}}, \bibinfo {author} {\bibfnamefont {K.~L.}\ \bibnamefont
  {Dooley}}, \bibinfo {author} {\bibfnamefont {H.}~\bibnamefont {Vahlbruch}},
  \bibinfo {author} {\bibfnamefont {C.}~\bibnamefont {Affeldt}}, \bibinfo
  {author} {\bibfnamefont {A.}~\bibnamefont {Bisht}}, \bibinfo {author}
  {\bibfnamefont {J.~R.}\ \bibnamefont {Leong}}, \bibinfo {author}
  {\bibfnamefont {J.}~\bibnamefont {Lough}}, \bibinfo {author} {\bibfnamefont
  {M.}~\bibnamefont {Prijatelj}}, \bibinfo {author} {\bibfnamefont
  {J.}~\bibnamefont {Slutsky}}, \bibinfo {author} {\bibfnamefont
  {M.}~\bibnamefont {Was}}, \bibinfo {author} {\bibfnamefont {H.}~\bibnamefont
  {Wittel}}, \bibinfo {author} {\bibfnamefont {K.}~\bibnamefont {Danzmann}},\
  and\ \bibinfo {author} {\bibfnamefont {H.}~\bibnamefont {Grote}},\ }\href
  {https://doi.org/10.1364/OE.24.000146} {\bibfield  {journal} {\bibinfo
  {journal} {Opt. Express}\ }\textbf {\bibinfo {volume} {24}},\ \bibinfo
  {pages} {146} (\bibinfo {year} {2016})}\BibitemShut {NoStop}%
\bibitem [{\citenamefont {Walker}\ \emph {et~al.}(2018)\citenamefont {Walker},
  \citenamefont {Agnew}, \citenamefont {Bidler}, \citenamefont {Lundgren},
  \citenamefont {Macedo}, \citenamefont {Macleod}, \citenamefont {Massinger},
  \citenamefont {Patane},\ and\ \citenamefont {Smith}}]{Walker2018}%
  \BibitemOpen
  \bibfield  {author} {\bibinfo {author} {\bibfnamefont {M.}~\bibnamefont
  {Walker}}, \bibinfo {author} {\bibfnamefont {A.~F.}\ \bibnamefont {Agnew}},
  \bibinfo {author} {\bibfnamefont {J.}~\bibnamefont {Bidler}}, \bibinfo
  {author} {\bibfnamefont {A.}~\bibnamefont {Lundgren}}, \bibinfo {author}
  {\bibfnamefont {A.}~\bibnamefont {Macedo}}, \bibinfo {author} {\bibfnamefont
  {D.}~\bibnamefont {Macleod}}, \bibinfo {author} {\bibfnamefont
  {T.}~\bibnamefont {Massinger}}, \bibinfo {author} {\bibfnamefont
  {O.}~\bibnamefont {Patane}},\ and\ \bibinfo {author} {\bibfnamefont {J.~R.}\
  \bibnamefont {Smith}},\ }\href@noop {} {\bibfield  {journal} {\bibinfo
  {journal} {Classical and Quantum Gravity}\ }\textbf {\bibinfo {volume}
  {35}},\ \bibinfo {pages} {225002} (\bibinfo {year} {2018})}\BibitemShut
  {NoStop}%
\bibitem [{\citenamefont {Collett}\ and\ \citenamefont
  {Gardiner}(1984)}]{Collett1984}%
  \BibitemOpen
  \bibfield  {author} {\bibinfo {author} {\bibfnamefont {M.~J.}\ \bibnamefont
  {Collett}}\ and\ \bibinfo {author} {\bibfnamefont {C.~W.}\ \bibnamefont
  {Gardiner}},\ }\href {https://doi.org/10.1103/PhysRevA.30.1386} {\bibfield
  {journal} {\bibinfo  {journal} {Phys. Rev. A}\ }\textbf {\bibinfo {volume}
  {30}},\ \bibinfo {pages} {1386} (\bibinfo {year} {1984})}\BibitemShut
  {NoStop}%
\bibitem [{\citenamefont {McCuller}\ \emph {et~al.}(2021)\citenamefont
  {McCuller}, \citenamefont {Dwyer}, \citenamefont {Green}, \citenamefont {Yu},
  \citenamefont {Kuns}, \citenamefont {Barsotti} \emph
  {et~al.}}]{McCuller2021}%
  \BibitemOpen
  \bibfield  {author} {\bibinfo {author} {\bibfnamefont {L.}~\bibnamefont
  {McCuller}}, \bibinfo {author} {\bibfnamefont {S.~E.}\ \bibnamefont {Dwyer}},
  \bibinfo {author} {\bibfnamefont {A.~C.}\ \bibnamefont {Green}}, \bibinfo
  {author} {\bibfnamefont {H.}~\bibnamefont {Yu}}, \bibinfo {author}
  {\bibfnamefont {K.}~\bibnamefont {Kuns}}, \bibinfo {author} {\bibfnamefont
  {L.}~\bibnamefont {Barsotti}}, \emph {et~al.},\ }\href
  {https://doi.org/10.1103/PhysRevD.104.062006} {\bibfield  {journal} {\bibinfo
   {journal} {Phys. Rev. D}\ }\textbf {\bibinfo {volume} {104}},\ \bibinfo
  {pages} {062006} (\bibinfo {year} {2021})}\BibitemShut {NoStop}%
\bibitem [{\citenamefont {Rahimi}\ and\ \citenamefont
  {Recht}(2007)}]{Rahimi2007}%
  \BibitemOpen
  \bibfield  {author} {\bibinfo {author} {\bibfnamefont {A.}~\bibnamefont
  {Rahimi}}\ and\ \bibinfo {author} {\bibfnamefont {B.}~\bibnamefont {Recht}},\
  }\href
  {https://papers.nips.cc/paper/2007/hash/013a006f03dbc5392effeb8f18fda755-Abstract.html}
  {\bibfield  {journal} {\bibinfo  {journal} {Advances in neural information
  processing systems}\ }\textbf {\bibinfo {volume} {20}} (\bibinfo {year}
  {2007})}\BibitemShut {NoStop}%
\bibitem [{\citenamefont {Tancik}\ \emph {et~al.}(2020)\citenamefont {Tancik},
  \citenamefont {Srinivasan}, \citenamefont {Mildenhall}, \citenamefont
  {Fridovich-Keil}, \citenamefont {Raghavan}, \citenamefont {Singhal},
  \citenamefont {Ramamoorthi}, \citenamefont {Barron},\ and\ \citenamefont
  {Ng}}]{Tancik2020}%
  \BibitemOpen
  \bibfield  {author} {\bibinfo {author} {\bibfnamefont {M.}~\bibnamefont
  {Tancik}}, \bibinfo {author} {\bibfnamefont {P.}~\bibnamefont {Srinivasan}},
  \bibinfo {author} {\bibfnamefont {B.}~\bibnamefont {Mildenhall}}, \bibinfo
  {author} {\bibfnamefont {S.}~\bibnamefont {Fridovich-Keil}}, \bibinfo
  {author} {\bibfnamefont {N.}~\bibnamefont {Raghavan}}, \bibinfo {author}
  {\bibfnamefont {U.}~\bibnamefont {Singhal}}, \bibinfo {author} {\bibfnamefont
  {R.}~\bibnamefont {Ramamoorthi}}, \bibinfo {author} {\bibfnamefont
  {J.}~\bibnamefont {Barron}},\ and\ \bibinfo {author} {\bibfnamefont
  {R.}~\bibnamefont {Ng}},\ }\href
  {https://proceedings.neurips.cc/paper/2020/hash/55053683268957697aa39fba6f231c68-Abstract.html}
  {\bibfield  {journal} {\bibinfo  {journal} {Advances in Neural Information
  Processing Systems}\ }\textbf {\bibinfo {volume} {33}},\ \bibinfo {pages}
  {7537} (\bibinfo {year} {2020})}\BibitemShut {NoStop}%
\bibitem [{\citenamefont {Lim}\ \emph {et~al.}(2021)\citenamefont {Lim},
  \citenamefont {Ar{\i}k}, \citenamefont {Loeff},\ and\ \citenamefont
  {Pfister}}]{Lim2021}%
  \BibitemOpen
  \bibfield  {author} {\bibinfo {author} {\bibfnamefont {B.}~\bibnamefont
  {Lim}}, \bibinfo {author} {\bibfnamefont {S.~{\"O}.}\ \bibnamefont
  {Ar{\i}k}}, \bibinfo {author} {\bibfnamefont {N.}~\bibnamefont {Loeff}},\
  and\ \bibinfo {author} {\bibfnamefont {T.}~\bibnamefont {Pfister}},\ }\href
  {https://doi.org/10.1016/j.ijforecast.2021.03.012} {\bibfield  {journal}
  {\bibinfo  {journal} {International Journal of Forecasting}\ }\textbf
  {\bibinfo {volume} {37}},\ \bibinfo {pages} {1748} (\bibinfo {year}
  {2021})}\BibitemShut {NoStop}%
\bibitem [{\citenamefont {Xue}\ \emph {et~al.}(2016)\citenamefont {Xue},
  \citenamefont {Zhang}, \citenamefont {Browne},\ and\ \citenamefont
  {Yao}}]{Xue16}%
  \BibitemOpen
  \bibfield  {author} {\bibinfo {author} {\bibfnamefont {B.}~\bibnamefont
  {Xue}}, \bibinfo {author} {\bibfnamefont {M.}~\bibnamefont {Zhang}}, \bibinfo
  {author} {\bibfnamefont {W.~N.}\ \bibnamefont {Browne}},\ and\ \bibinfo
  {author} {\bibfnamefont {X.}~\bibnamefont {Yao}},\ }\href
  {https://doi.org/10.1109/TEVC.2015.2504420} {\bibfield  {journal} {\bibinfo
  {journal} {IEEE Transactions on Evolutionary Computation}\ }\textbf {\bibinfo
  {volume} {20}},\ \bibinfo {pages} {606} (\bibinfo {year} {2016})}\BibitemShut
  {NoStop}%
\bibitem [{\citenamefont {Goda}\ \emph {et~al.}(2005)\citenamefont {Goda},
  \citenamefont {McKenzie}, \citenamefont {Mikhailov}, \citenamefont {Lam},
  \citenamefont {McClelland},\ and\ \citenamefont {Mavalvala}}]{Goda2005}%
  \BibitemOpen
  \bibfield  {author} {\bibinfo {author} {\bibfnamefont {K.}~\bibnamefont
  {Goda}}, \bibinfo {author} {\bibfnamefont {K.}~\bibnamefont {McKenzie}},
  \bibinfo {author} {\bibfnamefont {E.~E.}\ \bibnamefont {Mikhailov}}, \bibinfo
  {author} {\bibfnamefont {P.~K.}\ \bibnamefont {Lam}}, \bibinfo {author}
  {\bibfnamefont {D.~E.}\ \bibnamefont {McClelland}},\ and\ \bibinfo {author}
  {\bibfnamefont {N.}~\bibnamefont {Mavalvala}},\ }\href
  {https://doi.org/10.1103/PhysRevA.72.043819} {\bibfield  {journal} {\bibinfo
  {journal} {Phys. Rev. A}\ }\textbf {\bibinfo {volume} {72}},\ \bibinfo
  {pages} {043819} (\bibinfo {year} {2005})}\BibitemShut {NoStop}%
\bibitem [{\citenamefont {Sobol}(2001)}]{Sobol2001}%
  \BibitemOpen
  \bibfield  {author} {\bibinfo {author} {\bibfnamefont {I.~M.}\ \bibnamefont
  {Sobol}},\ }\href {https://doi.org/10.1016/S0378-4754(00)00270-6} {\bibfield
  {journal} {\bibinfo  {journal} {Mathematics and computers in simulation}\
  }\textbf {\bibinfo {volume} {55}},\ \bibinfo {pages} {271} (\bibinfo {year}
  {2001})}\BibitemShut {NoStop}%
\bibitem [{\citenamefont {Bellemare}\ \emph {et~al.}(2013)\citenamefont
  {Bellemare}, \citenamefont {Naddaf}, \citenamefont {Veness},\ and\
  \citenamefont {Bowling}}]{bellemare2013arcade}%
  \BibitemOpen
  \bibfield  {author} {\bibinfo {author} {\bibfnamefont {M.~G.}\ \bibnamefont
  {Bellemare}}, \bibinfo {author} {\bibfnamefont {Y.}~\bibnamefont {Naddaf}},
  \bibinfo {author} {\bibfnamefont {J.}~\bibnamefont {Veness}},\ and\ \bibinfo
  {author} {\bibfnamefont {M.}~\bibnamefont {Bowling}},\ }\href@noop {}
  {\bibfield  {journal} {\bibinfo  {journal} {Journal of Artificial
  Intelligence Research}\ }\textbf {\bibinfo {volume} {47}},\ \bibinfo {pages}
  {253} (\bibinfo {year} {2013})}\BibitemShut {NoStop}%
\bibitem [{\citenamefont {Mnih}\ \emph {et~al.}(2015)\citenamefont {Mnih},
  \citenamefont {Kavukcuoglu}, \citenamefont {Silver}, \citenamefont {Rusu}
  \emph {et~al.}}]{Mnih2015dqn}%
  \BibitemOpen
  \bibfield  {author} {\bibinfo {author} {\bibfnamefont {V.}~\bibnamefont
  {Mnih}}, \bibinfo {author} {\bibfnamefont {K.}~\bibnamefont {Kavukcuoglu}},
  \bibinfo {author} {\bibfnamefont {D.}~\bibnamefont {Silver}}, \bibinfo
  {author} {\bibfnamefont {A.~A.}\ \bibnamefont {Rusu}}, \emph {et~al.},\
  }\href {https://doi.org/10.1038/nature14236} {\bibfield  {journal} {\bibinfo
  {journal} {Nature}\ }\textbf {\bibinfo {volume} {518}},\ \bibinfo {pages}
  {529} (\bibinfo {year} {2015})}\BibitemShut {NoStop}%
\bibitem [{\citenamefont {Vinyals}\ \emph {et~al.}(2019)\citenamefont
  {Vinyals}, \citenamefont {Babuschkin}, \citenamefont {Czarnecki},
  \citenamefont {Mathieu} \emph {et~al.}}]{Vinyals2019AlphaStar}%
  \BibitemOpen
  \bibfield  {author} {\bibinfo {author} {\bibfnamefont {O.}~\bibnamefont
  {Vinyals}}, \bibinfo {author} {\bibfnamefont {I.}~\bibnamefont {Babuschkin}},
  \bibinfo {author} {\bibfnamefont {W.~M.}\ \bibnamefont {Czarnecki}}, \bibinfo
  {author} {\bibfnamefont {M.}~\bibnamefont {Mathieu}}, \emph {et~al.},\ }\href
  {https://doi.org/10.1038/s41586-019-1724-z} {\bibfield  {journal} {\bibinfo
  {journal} {Nature}\ }\textbf {\bibinfo {volume} {575}},\ \bibinfo {pages}
  {350} (\bibinfo {year} {2019})}\BibitemShut {NoStop}%
\bibitem [{\citenamefont {Berner}\ \emph {et~al.}(2019)\citenamefont {Berner},
  \citenamefont {Brockman}, \citenamefont {Chan}, \citenamefont {Cheung} \emph
  {et~al.}}]{berner2019dota}%
  \BibitemOpen
  \bibfield  {author} {\bibinfo {author} {\bibfnamefont {C.}~\bibnamefont
  {Berner}}, \bibinfo {author} {\bibfnamefont {G.}~\bibnamefont {Brockman}},
  \bibinfo {author} {\bibfnamefont {B.}~\bibnamefont {Chan}}, \bibinfo {author}
  {\bibfnamefont {V.}~\bibnamefont {Cheung}}, \emph {et~al.},\ }\href@noop {}
  {\bibfield  {journal} {\bibinfo  {journal} {arXiv preprint arXiv:1912.06680}\
  } (\bibinfo {year} {2019})}\BibitemShut {NoStop}%
\bibitem [{\citenamefont {Fu}\ \emph {et~al.}(2021)\citenamefont {Fu},
  \citenamefont {Yang}, \citenamefont {Agrawal},\ and\ \citenamefont
  {Jaakkola}}]{fu2021tia}%
  \BibitemOpen
  \bibfield  {author} {\bibinfo {author} {\bibfnamefont {X.}~\bibnamefont
  {Fu}}, \bibinfo {author} {\bibfnamefont {G.}~\bibnamefont {Yang}}, \bibinfo
  {author} {\bibfnamefont {P.}~\bibnamefont {Agrawal}},\ and\ \bibinfo {author}
  {\bibfnamefont {T.}~\bibnamefont {Jaakkola}},\ }in\ \href@noop {} {\emph
  {\bibinfo {booktitle} {Proceedings of the 38th International Conference on
  Machine Learning}}},\ \bibinfo {series} {Proceedings of Machine Learning
  Research}, Vol.\ \bibinfo {volume} {139},\ \bibinfo {editor} {edited by\
  \bibinfo {editor} {\bibfnamefont {M.}~\bibnamefont {Meila}}\ and\ \bibinfo
  {editor} {\bibfnamefont {T.}~\bibnamefont {Zhang}}}\ (\bibinfo  {publisher}
  {PMLR},\ \bibinfo {year} {2021})\ pp.\ \bibinfo {pages}
  {3480--3491}\BibitemShut {NoStop}%
\bibitem [{\citenamefont {Kumar}\ \emph {et~al.}(2021)\citenamefont {Kumar},
  \citenamefont {Fu}, \citenamefont {Pathak},\ and\ \citenamefont
  {Malik}}]{Kumar2021RMA}%
  \BibitemOpen
  \bibfield  {author} {\bibinfo {author} {\bibfnamefont {A.}~\bibnamefont
  {Kumar}}, \bibinfo {author} {\bibfnamefont {Z.}~\bibnamefont {Fu}}, \bibinfo
  {author} {\bibfnamefont {D.}~\bibnamefont {Pathak}},\ and\ \bibinfo {author}
  {\bibfnamefont {J.}~\bibnamefont {Malik}},\ }in\ \href
  {http://arxiv.org/abs/2107.04034} {\emph {\bibinfo {booktitle} {Robotics
  Science and Systems}}}\ (\bibinfo {year} {2021})\BibitemShut {NoStop}%
\bibitem [{\citenamefont {Lee}\ \emph {et~al.}(2020)\citenamefont {Lee},
  \citenamefont {Hwangbo}, \citenamefont {Wellhausen}, \citenamefont {Koltun},\
  and\ \citenamefont {Hutter}}]{lee2020quadrupedal}%
  \BibitemOpen
  \bibfield  {author} {\bibinfo {author} {\bibfnamefont {J.}~\bibnamefont
  {Lee}}, \bibinfo {author} {\bibfnamefont {J.}~\bibnamefont {Hwangbo}},
  \bibinfo {author} {\bibfnamefont {L.}~\bibnamefont {Wellhausen}}, \bibinfo
  {author} {\bibfnamefont {V.}~\bibnamefont {Koltun}},\ and\ \bibinfo {author}
  {\bibfnamefont {M.}~\bibnamefont {Hutter}},\ }\href
  {https://doi.org/10.1126/scirobotics.abc5986} {\bibfield  {journal} {\bibinfo
   {journal} {Science Robotics}\ }\textbf {\bibinfo {volume} {5}},\ \bibinfo
  {pages} {eabc5986} (\bibinfo {year} {2020})}\BibitemShut {NoStop}%
\bibitem [{\citenamefont {Gunny}\ \emph {et~al.}(2022)\citenamefont {Gunny},
  \citenamefont {Rankin}, \citenamefont {Krupa}, \citenamefont {Saleem},
  \citenamefont {Nguyen}, \citenamefont {Coughlin}, \citenamefont {Harris},
  \citenamefont {Katsavounidis},\ and\ \citenamefont {Timm}}]{ErikNature}%
  \BibitemOpen
  \bibfield  {author} {\bibinfo {author} {\bibfnamefont {A.}~\bibnamefont
  {Gunny}}, \bibinfo {author} {\bibfnamefont {D.}~\bibnamefont {Rankin}},
  \bibinfo {author} {\bibfnamefont {J.}~\bibnamefont {Krupa}}, \bibinfo
  {author} {\bibfnamefont {M.}~\bibnamefont {Saleem}}, \bibinfo {author}
  {\bibfnamefont {T.}~\bibnamefont {Nguyen}}, \bibinfo {author} {\bibfnamefont
  {M.}~\bibnamefont {Coughlin}}, \bibinfo {author} {\bibfnamefont
  {P.}~\bibnamefont {Harris}}, \bibinfo {author} {\bibfnamefont
  {E.}~\bibnamefont {Katsavounidis}},\ and\ \bibinfo {author} {\bibfnamefont
  {B.}~\bibnamefont {Timm}, \bibfnamefont {nd~Holzman}},\ }\href
  {https://doi.org/10.1038/s41550-022-01651-w} {\bibfield  {journal} {\bibinfo
  {journal} {Nature Astronomy}\ }\textbf {\bibinfo {volume} {33}},\ \bibinfo
  {pages} {075009} (\bibinfo {year} {2022})}\BibitemShut {NoStop}%
\end{thebibliography}%
\bibliographystyle{apsrev4-2.bst}

\end{document}